\documentclass[english,10pt,letterpaper,abstracton]{scrartcl}
\pdfoutput=1
\usepackage[square,numbers,comma,sort&compress]{natbib}
\usepackage{algorithm2e}
\usepackage{amsmath}
\usepackage{caption}
\usepackage{mdwtab}
\usepackage{multirow}
\usepackage{upgreek}
\usepackage{amssymb}
\usepackage{color}
\usepackage{fullpage}
\usepackage{authblk}
\usepackage[T1]{fontenc}
\usepackage{lmodern}
\usepackage{graphicx}
\usepackage{mathtools}
\usepackage{microtype}
\usepackage{paralist}
\usepackage[listofformat=parens,subrefformat=parens]{subfig}
\usepackage{stmaryrd}
\usepackage{xspace}
\usepackage{wrapfig}
\usepackage{floatrow}
\usepackage{setspace}
\usepackage{enumitem}
\usepackage{hyperref}

\title{Automatic Completion of\\Distributed Protocols with
  Symmetry}
\author[$\dag$]{\large Rajeev Alur}
\author[$\dag$]{\large Mukund Raghothaman}
\author[$\dag\ddag$]{\large Christos Stergiou}
\author[$\ddag\sharp$]{\large\\Stavros Tripakis}
\author[$\dag$]{\large Abhishek Udupa}
\affil[$\dag$]{\large University of Pennsylvania, Philadelphia, USA}
\affil[$\ddag$]{\large University of California, Berkeley, USA}
\affil[$\sharp$]{\large Aalto University, Helsinki, Finland}
\date{\vspace*{-4ex}}
\newcommand{\bfpara}[2]{\vspace*{1mm}\noindent\textbf{#1}\ {#2}}

\newcommand{\compose}{\mid}

\newcommand{\true}{\ensuremath{\mathsf{true}}\xspace}
\newcommand{\false}{\ensuremath{\mathsf{false}}\xspace}
\newcommand{\guard}{\ensuremath{\mathsf{guard}}\xspace}
\newcommand{\update}{\ensuremath{\mathsf{update}}\xspace}
\newcommand{\id}{\texttt{id}\xspace}

\newcommand{\interpretation}{R\xspace}

\newcommand{\lhs}{\mbox{\emph{lhs}}\xspace}
\newcommand{\location}{l\xspace}

\newcommand{\myid}{\texttt{Pm}\xspace}
\newcommand{\otherid}{\texttt{Po}\xspace}
\newcommand{\outputchannel}{y\xspace}
\newcommand{\outputchannels}{O\xspace}

\newcommand{\rhs}{\mbox{\emph{rhs}}\xspace}
\newcommand{\red}[1]{\noindent{\color{red} #1}}

\newcommand{\statevaluation}{\sigma\xspace}
\newcommand{\statevaluations}{\Sigma\xspace}

\newcommand{\unknownfunctions}{U\xspace}
\newcommand{\up}{\mbox{\emph{up}}\xspace}
\newcommand{\x}{\mbox{\emph{x}}\xspace}
\newcommand{\efsmsk}{\mbox{\textsc{esm-s}}\xspace}

\newcommand{\permset}[1]{\ensuremath{\mathsf{perm}(#1)}\xspace}
\newcommand{\guardcmd}[1]{\ensuremath{\mathsf{guard}(#1)}\xspace}
\newcommand{\updatecmd}[1]{\ensuremath{\mathsf{update}(#1)}\xspace}
\newcommand{\cmd}{\ensuremath{\mathsf{cmd}}\xspace}
\newcommand{\weakestpre}[2]{\ensuremath{\mathsf{wp}}(#1, #2)\xspace}
\newcommand{\weakestprecmd}[2]{\ensuremath{\mathsf{wpcmd}}(#1, #2)\xspace}
\newcommand{\sygus}{{\small{\textsf{SyGuS}}}\xspace}
\newcommand{\zug}[1]{\ensuremath{\langle#1\rangle}\xspace}
\newcommand{\ie}{\emph{i.e.}\xspace}
\newcommand{\pxrightarrow}[1]{\mathrel{\raisebox{-1.75pt}{$\xrightarrow{#1}$}}}

\usepackage{amsmath}
\usepackage{mathtools}
\usepackage{tikz}
\usepackage{xspace}
\usetikzlibrary{arrows}
\usetikzlibrary{automata}
\usetikzlibrary{calc}
\usetikzlibrary{decorations.pathmorphing}
\usetikzlibrary{positioning}
\usetikzlibrary{shapes}
\usetikzlibrary{shapes.geometric}
\usetikzlibrary{shapes.misc}
\usetikzlibrary{decorations}
\usetikzlibrary{automata, arrows, positioning}
\tikzset{>=stealth', state/.append style={minimum size=.4cm, align=center}}
\definecolor{dark-green}{rgb}{0,.6,0}

\tikzstyle{transition}=[->, >=stealth]
\tikzstyle{execution state}=[rectangle, draw, text centered, rounded corners, align=center] % text width minimum height=4em
\tikzstyle{transition label}=[above, align=center]

\usepackage{amsmath}
\usepackage{tikz}
\usetikzlibrary{arrows}
\usetikzlibrary{automata}
\usetikzlibrary{calc}
\usetikzlibrary{decorations.pathmorphing}
\usetikzlibrary{positioning}
\usetikzlibrary{shapes}
\usetikzlibrary{shapes.geometric}
\usetikzlibrary{shapes.misc}
\usetikzlibrary{decorations}
\usetikzlibrary{automata, arrows, positioning}
\tikzset{>=stealth', state/.append style={minimum size=.4cm}}
\definecolor{dark-green}{rgb}{0,.6,0}
\tikzstyle{transition}=[->, >=stealth]
\tikzstyle{transition label}=[above, align=center]

\usetikzlibrary{shapes.geometric,backgrounds,calc}

\tikzstyle{block} = [rectangle, draw, fill=blue!20, text centered, rounded corners]
\tikzstyle{line} = [draw, -latex']
\tikzstyle{cloud} = [draw, ellipse,fill=red!20, node distance=3cm,
    minimum height=2em]

\tikzset{
  basic box/.style = {
    shape = rectangle,
    align = center,
    draw  = black,
    font=\small,
    fill  = #1!25,
    inner sep = 3mm,
    thick,
    rounded corners},
  input box/.style = {
    shape = rectangle,
    align = center,
    draw  = black,
    font=\small,
    fill  = black!10,
    inner sep = 3mm,
    very thick},
  decision/.style = {
    shape = diamond,
    align = center,
    draw  = black,
    fill  = #1!25,
    text badly centered,
    inner sep=1mm,
    font=\small,
    thick
  },
  header node/.style = {
    Minimum Width = header nodes,
    font          = \strut\Large\ttfamily,
    text depth    = +0pt,
    font=\small,
    fill          = white,
    draw},
  header/.style = {%
    inner ysep = +1.5em,
    append after command = {
      \pgfextra{\let\TikZlastnode\tikzlastnode}
      node [header node] (header-\TikZlastnode) at (\TikZlastnode.north) {#1}
      node [span = (\TikZlastnode) (header-\TikZlastnode)]
        at (fit bounding box) (h-\TikZlastnode) {}
    }
  },
  hv/.style = {to path = {-|(\tikztotarget)\tikztonodes}},
  vh/.style = {to path = {|-(\tikztotarget)\tikztonodes}},
  fat blue line/.style = {ultra thick, blue},
  data line/.style = {ultra thick, black, dotted}
}

\DeclareCaptionFormat{ruledcapformat}{#1#2#3\hrulefill}
\begin{document}
\begin{spacing}{1.05}

\maketitle

\begin{abstract}
\noindent
A distributed protocol is typically modeled as a set of communicating processes,
where each process is described as an extended state machine along with fairness
assumptions, and its correctness is specified using safety and liveness requirements.
Designing correct distributed protocols is a challenging task. Aimed at simplifying this task,
we allow the designer to leave some of the guards and updates to state variables
in the description of extended state machines as unknown functions. The protocol
completion problem then is to find interpretations for these unknown functions while
guaranteeing correctness. In many distributed protocols, process behaviors are naturally
symmetric, and thus, synthesized expressions are further required to obey symmetry constraints.
Our counterexample-guided synthesis algorithm consists of repeatedly invoking two phases.
In the first phase, candidates for unknown expressions are generated using the SMT solver Z3.
This phase requires carefully orchestrating constraints to enforce the desired symmetry
in read/write accesses. In the second phase, the resulting completed protocol is checked
for correctness using a custom-built model checker that handles fairness assumptions, safety and
liveness requirements, and exploits symmetry. When model checking fails, our tool examines
a set of counterexamples to safety/liveness properties to generate constraints on unknown functions
that must be satisfied  by subsequent completions.  For evaluation, we show that our prototype
is able to automatically discover interesting missing details in distributed protocols
for mutual exclusion, self stabilization, and cache coherence.
\end{abstract}

\section{Introduction}
Protocols for coordination among concurrent processes are an essential
component of modern multiprocessor and distributed systems.  The
multitude of behaviors arising due to asynchrony and concurrency makes
the design of such protocols difficult. Consequently, analyzing such
protocols has been a central theme of research in formal verification
for decades. Now that verification tools are mature enough to be
applied to find bugs in real-world protocols, a promising area of
research is \emph{protocol synthesis}, aimed at simplifying the design
process via more intuitive programming abstractions to specify the
desired behavior.

Traditionally, a distributed protocol is modeled as a set of
communicating processes, where each process is described by an
extended state machine.  The correctness is specified by both safety
and liveness requirements.  In {\em reactive
synthesis}~\cite{RW89,PnueliRosner,BJPPS12}, the goal is to
automatically derive a protocol from its correctness requirements
specified in temporal logic.  However, if we require the
implementation to be distributed, then reactive synthesis is
undecidable~\cite{PR90,LamouchiThistle00,TripakisIPL04,FS05}.  An
alternative, and potentially more feasible approach inspired by {\em
program sketching}~\cite{sketch}, is to ask the programmer to specify
the protocol as a set of communicating state machines, but allow some
of the guards and updates to state variables to be unknown functions, to
be completed by the synthesizer so as to satisfy all the correctness
requirements.  This methodology for protocol specification can be
viewed as a fruitful collaboration between the designer and the
synthesis tool: the programmer has to describe the structure of the
desired protocol, but some details that the programmer is unsure
about, for instance, regarding corner cases and handling of unexpected
messages, are filled in automatically by the tool.

In our formalization of the synthesis problem, processes communicate
using input/output channels that carry typed messages. Each
process is described by a state machine with a set of typed
state variables.  Transitions consist of guards that test input
messages and state variables and updates to state variables and
fields of messages to be sent.  Such guards and updates can involve
{\em unknown\/} (typed) functions to be filled in by the synthesizer.
In many distributed protocols, such as cache coherence protocols,
processes are expected to behave in a symmetric manner. Thus,
we allow variables to have \emph{symmetric types} that restrict the
read/write accesses to obey symmetry constraints.  To specify safety
and liveness requirements, the state machines can be augmented with
acceptance conditions that capture incorrect executions. Finally,
fairness assumptions are added to restrict incorrect executions to
those that are \emph{fair}. It is worth noting that in
verification one can get useful analysis results by focusing solely on
safety requirements. In synthesis, however, ignoring liveness requirements and
fairness assumptions, typically results in trivial solutions.  The
protocol completion problem, then, is, given a set of extended state
machines with unknown guards and update functions, to find expressions
for the unknown functions so that the composition of the
resulting machines does not have an accepting fair execution.

Our synthesis algorithm relies on a counterexample-guided strategy
with two interacting phases: candidate interpretations for unknown
functions are generated using the SMT solver Z3 and the resulting
completed protocol is verified using a model checker.  We believe that
our realization of this strategy leads to the following contributions.
First, while searching for candidate interpretations for unknown
functions, we need to generate constraints that enforce symmetry in an
accurate manner without choking current SMT solvers.  Second,
surprisingly there is no publicly available model checker that handles
all the features that we critically need, namely, symmetry, liveness
requirements, and fairness assumptions.  As a result, building on the
known theoretical foundations, we had to develop our own model checker
(which we plan to make publicly available).  Third, we develop an
algorithm that examines the counterexamples to safety/liveness
requirements when model checking fails, and generates constraints on
unknown functions that must be satisfied in subsequent completions.
Finally, the huge search space for candidate expressions is a
challenge for the scalability for any synthesis approach.  As reported in
section 4, we experimented with many alternative strategies for
prioritizing the search for candidate expressions, and this experience
offers some insights regarding what information a user can provide for
getting useful results from the synthesis tool. We evaluate our
synthesis tool in completing a mutual exclusion protocol, a self
stabilization protocol and a non-trivial cache coherence
protocol. Large parts of the behavior of the protocol were left
unspecified in the case of the mutual exclusion protocol and the self
stabilization protocol, whereas the cache coherence protocol had quite
a few tricky details left unspecified. Our tool was able
to synthesize the correct completions for all these protocols in a
reasonable amount of time.

\bfpara{Related Work.}{\emph{Bounded synthesis}~\cite{FS13} and
    \emph{genetic programming}~\cite{KP08,KatzPeled09} are other
    approaches for handling the undecidability of distributed reactive
    synthesis.  In the first, the size of the implementation is
    restricted, thus allowing for algorithmic solutions, and in the
    second the implementation space is sampled and candidates are
    mutated in a stochastic process.  The problem of inferring
    extended finite-state machines has been studied in the context of
    active learning~\cite{learningefsm}.  The problem of completing
    distributed protocols has been targeted by the works presented in
    \cite{hvc,transit} and \emph{program repair}~\cite{JGB05}
    addresses a similar problem. Compared to \cite{hvc}, our algorithm
    can handle extended state machines that include variables and
    transitions with symbolic expressions as guards and updates.
    Compared to \cite{transit}, our algorithm can also handle liveness
    violations and, more importantly, can process counterexamples
    automatically and, thus, does not require a human in the synthesis
    loop.  \textsc{psketch}~\cite{psketch} is an extension of the
    \emph{program sketching} work for concurrent data structures but
    is limited to safety properties. The work in \cite{tiwari}
    describes an approach based on QBF solvers for synthesizing a
    distributed self-stabilizing system, which also approximates
    liveness with safety and uses templates for the synthesized
    functions.  Last, compared to all works mentioned above, our
    algorithm can be used to enforce symmetry in the synthesized
    processes.}

%%% Local Variables:
%%% TeX-master: "cav15"
%%% End:

\section{An Illustrative Example}
\label{sec:Motivation}
%!TEX root = cav15.tex

\begin{figure}[!t]
\centering
\subfloat[Parameterized Symmetric Process]{
  %!TEX root = ../../cav15.tex

\begin{tikzpicture}
  \path
  node[state, initial, initial text={}] (L1) {$L_1$}
  node[state, right=2.5cm of L1] (L2) {$L_2$}
  node[state, right=2.5cm of L2] (L3) {$L_3$}
  node[state, below=.5cm of L2] (Critical) {$L_4$}
  node[below=.03cm of Critical] {\textsf{critical section}}
  ;
  \path
  (L1)
  edge[transition]
  node[transition label]
  {
  $\text{request}_\myid{}!$\\
  $\text{flag}[\myid{}] \coloneqq\true$
  }
  (L2)
  (L2)
  edge[transition]
  node[transition label]
  {
  $\text{turn} \coloneqq\otherid{}$
  }
  (L3)
  (L3)
  edge[transition, out=70, in=110, looseness=6]
  node[transition label]
  {
  $\text{flag}[\otherid] \land\text{turn} = \otherid$\\
  $\text{waiting}_\myid!$
  }
  (L3)
  (L3)
  edge[transition]
  node[transition label, below right=.05cm and -.4cm]
  {$\lnot \text{flag}[\otherid] \lor \text{turn} = \myid$\\
  $\text{critical}_\myid!$}
  (Critical)
  (Critical)
  edge[transition]%, out=230, in=-50]
  node[transition label, below left=.1cm and -.6cm]
  {
  $\text{flag}[\myid]\coloneqq \false$
  }
  (L1)
  ;
\end{tikzpicture}
  \label{figure:peterson_skeleton}
}\\
\subfloat[Incomplete process sketch]{
  %!TEX root = ../../cav15.tex

\begin{tikzpicture}
  \path
  node[state, initial, initial text={}] (L1) {$L_1$}
  node[state, right=2.5cm of L1] (L2) {$L_2$}
  node[state, right=2.5cm of L2] (L3) {$L_3$}
  node[state, below=.5cm of L2] (Critical) {$L_4$}
  node[below=.03cm of Critical] {\textsf{critical section}}
  ;
  \path
  (L1)
  edge[transition]
  node[transition label]
  {
  $\text{request}_\myid!$\\
  $\text{flag}[\myid] \coloneqq \true$
  }
  (L2)
  (L2)
  edge[transition]
  node[transition label]
  {
  $\text{turn} \coloneqq \otherid$
  }
  (L3)
  (L3)
  edge[transition, out=70, in=110, looseness=6]
  node[transition label]
  {
  $g_\textrm{wait}(\myid, \otherid, \text{flag}, \text{turn})$\\
  $\text{waiting}_\myid!$
  }
  (L3)
  (L3)
  edge[transition]
  node[transition label, below right=.05cm and -.4cm]
  {
  $g_\textrm{crit}(\myid, \otherid, \text{flag}, \text{turn})$\\
  $\text{critical}_\myid!$}
  (Critical)
  (Critical)
  edge[transition]%, out=230, in=-50]
  node[transition label, below left=.1cm and -.6cm]
  {
  $\text{flag}[\myid]\coloneqq \false$
  }
  (L1)
  ;
\end{tikzpicture}
  \label{figure:peterson_sketch}
}\\
\subfloat[Liveness Monitor for Peterson's Algorithm]{
  %!TEX root = ../../cav15.tex

\begin{tikzpicture}
  \path
  node[state, initial, initial text={},inner sep=2pt] (L1) {$M_1$}
  node[state, right=2cm of L1, accepting,inner sep=2pt] (L2) {$M_2$}
  node[state, right=2cm of L2,inner sep=2pt] (L3) {$M_3$}
  ;
  \path
  (L1)
  edge[transition, loop above]
  node[transition label]
  {
  $\text{waiting}_\id?$\\
  $\text{request}_\id?$\\
  $\text{critical}_\id?$
  }
  (L1)
  (L1)
  edge[transition]
  node[transition label] {$\text{request}_\id?$}
  (L2)
  (L2)
  edge[transition, loop above]
  node[transition label]
  {
  $\text{waiting}_\id?$\\
  $\text{request}_\id?$
  }
  (L2)
  (L2)
  edge[transition]
  node[transition label] {$\text{critical}_\id?$}
  (L3)
  (L3)
  edge[transition, loop above]
  node[transition label]
  {
  $\text{waiting}_\id?$\\
  $\text{request}_\id?$\\
  $\text{critical}_\id?$
  }
  (L3)
  ;
\end{tikzpicture}
  \label{figure:peterson_liveness_monitor}
}
\caption{Peterson's mutual exclusion algorithm. Note that the non-trivial
  guards of the $(L_3, L_3)$ and $(L_3, L_4)$ transitions in
  Figure~\protect\subref*{figure:peterson_skeleton} have been replaced in
  Figure~\protect\subref*{figure:peterson_sketch} by ``unknown'' functions
  $g_{\mathrm{wait}}$ and $g_{\mathrm{crit}}$ respectively.}
\label{figure:peterson-figures}
\end{figure}
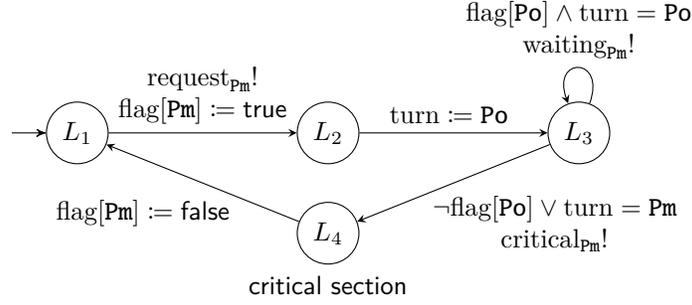
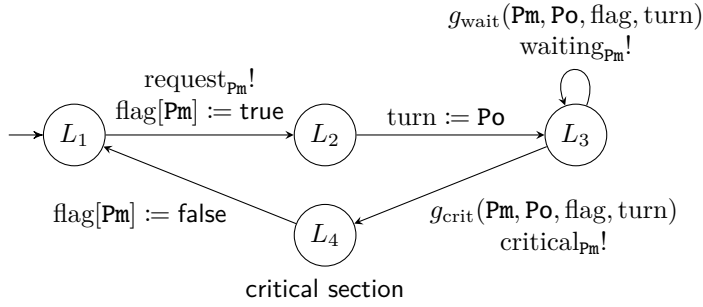
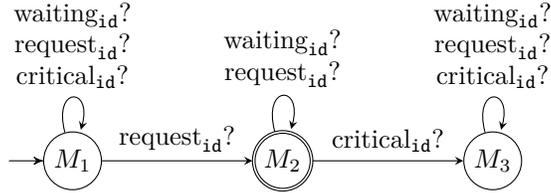

Consider Peterson's mutual exclusion algorithm, described in
Figure~\subref*{figure:peterson_skeleton}, which manages two symmetric
processes contending for access to a critical section. Each process
is parameterized by $\texttt{Pm}$ and $\texttt{Po}$ (for
``my'' process id and ``other'' process id respectively),
such that $\texttt{Pm} \neq \texttt{Po}$. Both parameters
$\texttt{Pm}$ and $\texttt{Po}$ are of type \textsf{processid}
and they are allowed to take on values \texttt{P0} and
\texttt{P1}. We therefore have two instances:
$P_0$, where $(\texttt{Pm} = \texttt{P0}, \texttt{Po} = \texttt{P1})$,
and $P_1$, where $(\texttt{Pm} = \texttt{P1}, \texttt{Po} = \texttt{P0})$.
%% We now describe a scenario where protocol completion can be useful,
%% using Peterson's mutual exclusion protocol as an illustrative example.
%% Figure~\ref{figure:peterson_skeleton} describes the working
%% of the two symmetric processes in Peterson's mutual exclusion
%% protocol. The state machine in Figure~\ref{figure:peterson_skeleton}
%% is parametrized by two parameters $\texttt{Pm}$ and $\texttt{Po}$ (for
%% ``my'' process id and ``other'' process id respectively), with the
%% constraint that $\texttt{Pm} \neq \texttt{Po}$. The type of both
%% $\texttt{Pm}$ and $\texttt{Po}$ is \textsf{processid}
%% and they are allowed to take on values \texttt{P0} and
%% \texttt{P1}. Instantiating the parameters thus gives us two processes,
%% one with $\texttt{Pm} = \texttt{P0}$, and $\texttt{Po} = \texttt{P1}$
%% which we call $P_0$ and the symmetric process which we call $P_1$.
$P_0$ and $P_1$ communicate through the shared variables \emph{turn} and
\emph{flag}. The variable \emph{turn} has type \textsf{processid}.
The \emph{flag} variable is an array of Boolean values, with index
type \textsf{processid}. The main objective of the protocol is to control
access to the critical section, represented by location $L_4$, and
ensure that both of the processes $P_0$ and $P_1$ are never
simultaneously in the critical section, \emph{i.e.}, it is a safety
violation for both $P_0$ and $P_1$ to be in state $L_4$ at the same time.

The liveness monitor shown in
Figure~\subref*{figure:peterson_liveness_monitor} captures the requirement
that a process does not wait indefinitely to enter the critical
section. The monitor accepts all undesirable runs where a process has requested
access to the critical section but never reaches state $L_4$
after. The messages $\text{request}$, $\text{waiting}$, and
$\text{critical}$ inform the liveness monitor about
the state of the processes, and the synchronization model here is that
of communicating I/O automata~\cite{lynch-book-96}. Note that a run
accepted by the monitor may be \emph{unfair} with respect to some
processes. Enforcing \emph{weak} process fairness on $P_0$ and $P_1$,
--- \emph{i.e.}, if a process is enabled at every point in an
accepting cycle, then it must be executed at some point in the cycle
--- is sufficient to rule out unfair executions, but not
necessary. Enforcing weak fairness on the transitions between $(L_2,
L_3)$, $(L_3, L_4)$ and $(L_4, L_1)$ suffices.

Now, suppose the protocol developer has trouble figuring out the
exact guard under which a process is allowed to enter the
critical section, but knows the structure of the
processes $P_0$ and $P_1$, and requires them to be
symmetric. Figure~\subref*{figure:peterson_sketch} describes what the
developer knows about the protocol. The functions $g_{\mathrm{wait}}$ and
$g_{\mathrm{crit}}$ represent unknown Boolean valued functions over
the state variables and the parameters of the process
under consideration.
%% TODO: Rephrase the next sentence?
Including the parameters as part of the domain of
$g_{\mathrm{wait}}$ and $g_{\mathrm{crit}}$ indicates that the
completions for processes $P_0$ and $P_1$ need to be symmetric.
The objective is to assist the developer by automatically discovering
interpretations for these unknown functions, such  that the completed
protocol satisfies the necessary mutual exclusion property, and the
requirements imposed by the liveness monitor. We formalize this
completion problem in Section~\ref{sec:Formalization}, and present our
completion algorithm in Section~\ref{sec:Completion}.

\section{Formalization}
\label{sec:Formalization}

%!TEX root = cav15.tex

%% TODO: use consistent font for "values"

%% TODO: Make symbols for types consistent: Uniformly capital $T$.

%% TODO: Fix strong and weak fairnesses in the definition of composition in the
%%   appendix.

\subsection{Extended State Machine Sketches}
\label{subsection:efsms}
\global\long\def\tybool{\textit{bool}}
\global\long\def\tynat{\textit{nat}}

We model processes using Extended State Machine Sketches ($\efsmsk$).
Fix a collection of types, such as the type $\tybool$ of the Boolean values
$\{ \true, \false \}$, enumerated types such as $\{\mathsf{red},
\mathsf{green}, \mathsf{blue}\}$, or finite subsets $\tynat[x, y]$ of natural
numbers $\{ i \mid x \leq i \leq y\}$. Other examples of types include
\emph{symmetric types} (described in Section~\ref{subsection:symmetry}), and
array types, which map each value of the index type to a value of the range
type. Note that the cardinality of each type is required to be finite.

The description of an $\efsmsk$ will mention several function symbols. Some
of these, such as the operator ``$+$'' for addition, have interpretations
which are already known, while others, such as the guard $g_{\mathrm{crit}}$
to enter the critical section in the incomplete sketch of Peterson's algorithm have
unknown interpretations. Each function symbol, both known and unknown, is
associated with a signature, $d_1 \times \cdots \times d_n \to r$, where
$d_1$, \ldots, $d_n$ are the types of its arguments and $r$ is the return type.
Expressions may then be constructed, such as ``$\textit{timestamp} + 1$'',
using these function symbols, state variables, and input channels.
Formally, an \efsmsk $A$ is a tuple $\langle L, \location_0, I, O, S,
  \statevaluation_{0}, \unknownfunctions, T, \mathcal{F}_s, \mathcal{F}_w
  \rangle$ such that:
\begin{itemize}[itemsep=1pt,topsep=3pt,parsep=3pt]
\item $L$ is a finite set of locations and $l_0 \in L$ is the initial location,
\item $I$ and $O$ are finite sets of typed input and output channels, respectively,
\item $S$ is a finite set of typed state variables,
\item $\statevaluation_{0}$ maps each variable $x \in S$ to its initial value
  $\statevaluation_{0} (x)$,
\item $\unknownfunctions$ is a set of unknown function symbols,
\item $T$ is a set of transitions of the form $\langle \location$, $c$,
  $\guard$, $\update$, $\location'\rangle$, where $c \in I$ for input, $c
  \in O$ for output, and $c = \epsilon$ for internal transitions, $\guard$
  is the transition guard, and $\update$ are the transition updates,
\item $\mathcal{F}_s, \mathcal{F}_w \subseteq 2^{T_\epsilon \cup
T_O}$, are sets of strong and weak fairnesses respectively. Here $T_O$
and $T_{\epsilon}$ are the sets of output and internal transitions
respectively.
\end{itemize}
A guard description $\guard$ is a Boolean expression over the state
variables $S$ that can use unknown functions from $U$. Similarly, an
update description $\update$ is a sequence of assignments of the form $\lhs
\coloneqq \rhs$ where $\lhs$ is one of the state variables or an output
channel in the case of an output transition, and $\rhs$ is an expression
over state variables or state variables and an input channel in the case of
an input transition, possibly using unknown functions from $U$.

\bfpara{Executions.}{
To define the executions of an $\efsmsk$, we first pick an \emph{interpretation}
$\interpretation$ which maps each unknown function $u \in U$ to an interpretation
of $u$. Given a set of variables $V$, a \emph{valuation} $\statevaluation$ is a
function which maps each variable $x \in V$ to a value $\statevaluation(x)$ of
the corresponding type, and we write $\statevaluations_V$ for the set of all
such valuations. Given a valuation $\statevaluation \in \statevaluations_V$,
a variable $x$, and a value $v$ of appropriate type, we write
$\statevaluation[x \mapsto v] \in \statevaluations_{V \cup \{ x \}}$ for the
valuation which maps all variables $y \neq x$ to $\statevaluation(y)$, and maps
$x$ to $v$.}

The executions of $A$ are defined by describing the updates to the
state valuation $\statevaluation \in \statevaluations_S$ during each
transition. Note that each guard description $\guard$ naturally
defines a set $\llbracket \guard, \interpretation \rrbracket$ of
valuations $\statevaluation \in \statevaluations_S$ which satisfy
$\guard$ with the unknown functions instantiated with
$\interpretation$.  Similarly, each update description $\update$
defines a function $\llbracket \update, \interpretation \rrbracket$ of
type $\statevaluations_{S \cup \{ x \}} \to \statevaluations_{S}$ for
input transitions on the channel $x$, $\statevaluations_{S} \to
\statevaluations_{S \cup \{ y \}}$ for output transitions on the
channel $y$, and $\statevaluations_{S} \to \statevaluations_{S}$ for
internal transitions respectively.

A \emph{state} of an \efsmsk $A$ is a pair $(\location, \statevaluation)$
of a location $l \in L$ and a state valuation $\statevaluation \in
  \statevaluations_S$. We then write:
\begin{itemize}[itemsep=1pt,topsep=3pt,parsep=3pt]
\item $(\location, \statevaluation) \pxrightarrow{x?v} (l', \statevaluation')$ if
  $A$ has an input transition from $\location$ to $\location'$ on channel
  $x$ with guard $\guard$ and update $\update$ such that
  $\statevaluation \in \llbracket \guard, \interpretation \rrbracket$ and
  $\llbracket \update, \interpretation \rrbracket(\statevaluation[x \mapsto v])
    = \statevaluation'$;
\item
$(l, \statevaluation) \pxrightarrow{y!v} (l', \statevaluation')$ if $A$ has
  an output transition from $\location$ to $\location'$ on channel $y$ with
  guard $\guard$ and update $\update$ such that $\statevaluation \in
  \llbracket \guard, \interpretation \rrbracket$ and $\llbracket \update,
  \interpretation \rrbracket(\statevaluation) = \statevaluation'[y \mapsto
    v]$; and
\item
$(l, \statevaluation) \pxrightarrow{\epsilon} (l', \statevaluation')$ if $A$
  has an internal transition from $\location$ to $\location'$ with guard
  $\guard$ and update $\guard$ such that $\statevaluation \in \llbracket
  \guard, \interpretation \rrbracket$ and $\llbracket \update,
  \interpretation \rrbracket(\statevaluation) = \statevaluation'$.
\end{itemize}
We write $(l, \statevaluation) \to (l', \statevaluation')$ if either there
are $x, v$ such that $(\location, \statevaluation) \pxrightarrow{x?v} (l',
\statevaluation')$, there are $y, v$ such that $(\location,
\statevaluation) \pxrightarrow{y!v} (l', \statevaluation')$, or $(\location,
\statevaluation) \pxrightarrow{\epsilon} (l', \statevaluation')$.
A finite (infinite) \emph{execution} of the \efsmsk $A$ under
$\interpretation$ is then a finite (resp. infinite) sequence: $(\location_0,
\statevaluation_0) \to (\location_1, \statevaluation_1) \to (\location_2,
\statevaluation_2) \to \cdots$ where for every $j \ge 0$, $(l_j,
\statevaluation_j)$ is a state of $A$, $(\location_0, \statevaluation_0)$
is an initial state of $A$, and for $j \ge 1$, $(\location_j,
\statevaluation_j) \to (\location_{j+1}, \statevaluation_{j+1})$.
A state $(\location, \statevaluation)$ is \emph{reachable} under $R$ if there
exists a finite execution that reaches that state: $(\location_0,
\statevaluation_0) \to \cdots \to (\location, \statevaluation)$.
We say that a transition from $\location$ to $\location'$ with guard
$\guard$ is \emph{enabled} in state $(\location, \sigma)$ if $\sigma \in
\llbracket \guard, \interpretation \rrbracket$.  A state $(\location,
\sigma)$ is called a \emph{deadlock} if no transition is enabled in
$(\location, \sigma)$. The \efsmsk $A$ is called deadlock-free under $R$ if no
deadlock state is reachable under $R$.
The $\efsmsk$ $A$ is called \emph{deterministic} under $\interpretation$ if for
every state $(\location, \statevaluation)$, if there are multiple
transitions enabled at $(\location, \sigma)$, then they must be input
transitions on distinct input channels.

Consider a weak fairness requirement $F \in \mathcal{F}_w$.
An infinite execution of $A$ under $\interpretation$ is called \emph{fair} with
respect to a weak fairness $F$ if either:
\begin{inparaenum}[(\itshape a\upshape)]
  \item for infinitely many indices $i$, none of the transitions $t \in F$ is
    enabled in $(\location_i, \statevaluation_i)$, or
  \item for infinitely many indices $j$ one of the transitions in $F$ is taken
    at step $j$.
\end{inparaenum}
Thus, for example, the necessary fairness assumptions for Peterson's
algorithm are $\mathcal{F}_w = \{ \{ \tau_{23} \}, \{ \tau_{34} \}, \{ \tau_{41} \} \}$,
where $\tau_{23}$, $\tau_{34}$, and $\tau_{41}$ refer to the
$(L_2, L_3)$, $(L_3, L_4)$ and $(L_4, L_1)$ transitions respectively.
Similarly, an infinite execution of
$A$ under $\interpretation$ is fair with respect to a strong fairness
$F \in \mathcal{F}_s$ if either:
\begin{inparaenum}[(\itshape a\upshape)]
  \item there exists $k$ such that for every $i \ge k$ and every transition $t
      \in F$, $t$ is not enabled in $(\location_i, \statevaluation_i)$, or
  \item for infinitely many indices $j$ one of the transitions in $F$ is taken
    at step $j$.
\end{inparaenum}
Finally, an infinite execution of $A$ is fair if it is fair with respect to
each strong and weak fairness requirement in $\mathcal{F}_s$ and $\mathcal{F}_w$
respectively.

\bfpara{Composition of {\normalfont\scshape esm-s}}{
Informally, two
\efsmsk $A_1$ and $A_2$ are composed by synchronizing their output and
input transitions on a given channel. If $A_1$ has an output transition on
channel $c$ from location $l_1$ to $l_1'$ with guard and updates $\guard_1$
and $\update_1$, and $A_2$ has an input transition on the same channel $c$
from location $l_2$ to $l_2'$ with guard and updates $\guard_2$ and
$\update_2$ then their product has an output transition from location
$(l_1, l_2)$ to $(l_1', l_2')$ on channel $c$ with guard $\guard_1 \land
\guard_2$ and updates $\update_1 ; \update_2$.  Note that by sequencing the
updates, the value written to the channel $c$ by $A_1$ is then used by
subsequent updates of the variables of $A_2$ in $\update_2$. We now
provide a formal definition of the composition of two \textsc{esm} sketches.

Consider two $\efsmsk$ $A_1$ and $A_2$, where $A_1$ is of the form
$A_1 = \langle L_1, l_{0, 1}, I_1, O_1, S_1, \statevaluation_{0, 1},
\unknownfunctions_1, T_1, \mathcal{F}_{s1}, \mathcal{F}_{w1} \rangle$
and $A_2$ is of the form $A_2 = \langle L_2, l_{0, 2}, I_2, O_2, S_2,
\statevaluation_{0, 2}, \unknownfunctions_2, T_2, \mathcal{F}_{s2},
\mathcal{F}_{w2} \rangle$ such that $O_1 \cap O_2 = \emptyset$ and
$S_1 \cap S_2 = \emptyset$.  We then define their composition, $A_1
\compose A_2$, as the $\efsmsk$: $\langle L$, $\location_{0}$, $I$,
$O$, $S$, $\statevaluation_{0}$, $\unknownfunctions$, $T$,
$\mathcal{F}_s$, $\mathcal{F}_w$ $\rangle$ where:
\begin{itemize}[itemsep=0.75pt,topsep=1pt,parsep=1pt]
\item
    $L = L_1 \times L_2$,
\item
    $l_0 = \langle l_{0, 1}, l_{0, 2} \rangle$,
\item
	$I = (I_1 \cup I_2) \setminus (O_1 \cup O_2)$,
\item
	$O = O_1 \cup O_2$,
\item
	$S = S_1 \cup S_2$,
\item
	$\statevaluation_0 = \statevaluation_{0,1} \cup \statevaluation_{0,2}$,
\item
    $\unknownfunctions = \unknownfunctions_1 \cup \unknownfunctions_2$,
\item
    For every input channel $x \in I$,
    $\langle(\location_1, \location_2), x, \guard, \update,
    (\location_1', \location_2') \rangle \in T$ if and only if \emph{at
      least} one of the following holds:
    \begin{enumerate}[label=(\alph*)]
    \item
        $x \notin I_2$,
        $\location_2 = \location_2'$, and
        $\langle \location_1, x,\guard, \update, \location_1' \rangle \in T_1$,
    \item
        $x \notin I_1$,
        $\location_1 = \location_1'$, and
        $\langle \location_2, x, \guard, \update, \location_2' \rangle \in T_2$,
    \item
        $x \in I_1\cap I_2$,
        $\langle \location_1, x, \guard_1, \update_1, \location_1'
        \rangle \in T_1$, $\langle \location_2, \guard_1,
        \update_1, \location_2' \rangle \in T_2$, $\guard =
        \guard_1 \land \guard_2$, and
        $\update  = \update_1 ; \update_2$,
    \end{enumerate}
\item
    For every output channel $\outputchannel \in \outputchannels$,
    $\langle(\location_1, \location_2), \outputchannel, \guard,
    \update,(\location_1', \location_2') \rangle \in T$ if and only if
    \emph{at least} one of the following holds:
    \begin{enumerate}[label=(\alph*)]
    \item
        $y \in O_1$,
        $y \notin I_2$,
        $\location_2 = \location_2'$, and
        $\langle \location_1, \outputchannel, \guard, \update, \location_1' \rangle \in T_1$,
    \item
        $y \in O_2$,
        $y \notin I_1$,
        $\location_1 = \location_1'$, and
        $\langle \location_2, \outputchannel, \guard, \update, \location_2' \rangle \in T_2$,
    \item
        $y \in O_1$,
        $y \in I_2$,
        $\langle \location_1, \guard_1, \update_1, \location_1' \rangle \in T_{1}$,
        $\langle \location_2, \guard_2, \update_2, \location_2' \rangle \in T_{2}$,
        $\guard = \guard_1 \land \guard_2$, and
        $\update = \update_1; \update_2$,
    \item
        $y \in O_2$,
        $y \in I_1$,
        $\langle \location_1, \outputchannel, \guard_1, \update_1, \location_1' \rangle \in T_1$,
        $\langle \location_2, \outputchannel, \guard_2, \update_2, \location_2' \rangle \in T_2$,
        $\guard = \guard_1 \land \guard_2$, and
        $\update = \update_2; \update_1$,
      \end{enumerate}
\item
    $\langle (\location_1, \location_2), \epsilon, \guard, \update,
    (\location_1', \location_2') \rangle \in T$ if and only if
    \emph{at least} one of the following hold:
    \begin{enumerate}[label=(\alph*)]
      \item
        $\langle \location_1, \epsilon, \guard, \update, \location_1'
        \rangle \in T_1$ and $\location_2 = \location_2'$.
      \item
        $\langle \location_2, \epsilon, \guard, \update, \location_2'
        \rangle \in T_2$, and $\location_1 = \location_1'$.
    \end{enumerate}
\item
    $\mathcal{F}_s = \{F^1, \ldots, F^N\}$ such that
    for every $F^i \in \mathcal{F}_s$, either:
    \begin{enumerate}[label=(\alph*)]
    \item
    there exists $F^j_1 \in \mathcal{F}_{s1}$ such that for every
    transition $t \in T$, $t \in F^i$ if and only if there exists a
    transition of the form $\langle \location_1, c, \guard_1, \update_1,
    \location_1' \rangle \in F^j_1$ and $t$ is of the form $\langle
    (\location_1, \location_2)$, $c$, $\guard_1 \land \guard_2$,
    $\update_1; \update_2$, $(\location_1', \location_2') \rangle$ or of
    the form $\langle (\location_1, \location_2)$, $c$, $\guard_1$,
    $\update_1$, $(\location_1', \location_2) \rangle$.
    \item
    there exists $F^j_2 \in \mathcal{F}_{s2}$ such that for every
    transition $t \in T$, $t \in F^i$ if and only if there exists a
    transition of the form $\langle \location_2, c, \guard_2, \update_2,
    \location_2' \rangle \in F^j_2$ and $t$ is of the form $\langle
    (\location_1, \location_2)$, $c$, $\guard_1 \land \guard_2$,
    $\update_1; \update_2$, $(\location_1', \location_2') \rangle$, or of
    the form $\langle (\location_1, \location_2)$, $c$, $\guard_2$,
    $\update_2$, $(\location_1, \location_2') \rangle$.
  \end{enumerate}
\item
    $\mathcal{F}_w$ is defined in the same way as $\mathcal{F}_s$
\end{itemize}
Note that the composition operator ``$\compose$'' is commutative and associative.
}

\bfpara{Specifications.}{An \efsmsk can be equipped with
  error locations $L_e \subseteq L$, accepting locations
  $L_a \subseteq L$, or both.\footnote{The error and accepting
    locations of an \efsmsk are designated in the figures using double
    circles.   Traditionally, \emph{safety and liveness monitors} have been used to
    characterize erroneous finite and infinite executions. In this
    spirit, if an \efsmsk is used solely for labeling product locations
    as error or accepting, we will call it a monitor, but still refer to
    the safety and liveness of the product \efsmsk.}
  The composition of two \efsmsk $A_1, A_2$ ``inherits''
  the error and accepting locations of its components.  A product
  location $(l_1, l_2)$ is an error (accepting) location if either
  $l_1$ or $l_2$ are error (accepting) locations.
  An \efsmsk $A$ is called \emph{safe} under $R$ if for all reachable
  states $(l, \statevaluation)$, $l$ is not an error location.  An
  infinite execution of $A$ under $R$,
  $(\location_0, \statevaluation_0) \to (\location_1,
  \statevaluation_1) \to \cdots$,
  is called \emph{accepting} if for infinitely many indices $j$,
  $\location_j \in L_a$.  $A$ is called \emph{live} under $R$ if it
  has no infinite fair accepting executions.
}

\subsection{Symmetry}
\label{subsection:symmetry}
It is often required that the processes of an $\efsmsk$ completion problem have
some structurally similar behavior, as we saw in Section~\ref{sec:Motivation}
in the case of Peterson's algorithm. To describe such requirements, we
use \emph{symmetric types}, which are similar to \emph{scalarsets} used in
the Mur$\varphi$ model checker~\cite{dill-96a}.

A symmetric type $T$ is characterized by:
\begin{inparaenum}[(\itshape a\upshape)]
  \item its name, and
  \item its cardinality $|T|$, which is a finite number.
\end{inparaenum}
Given a collection of processes
parameterized by a symmetric type $T$, such as $P_0$ and $P_1$ of Peterson's
algorithm, the idea is that the system is invariant under permutations (i.e.
renaming) of the parameter values.

Let $\permset{T}$ be the set of all permutations $\pi_T : T \to T$ over the
symmetric type $T$. For ease of notation, we define $\pi_T(v) = v$, for values
$v$ whose type is \emph{not} $T$. Given the collection of all symmetric
types $\mathcal{T} = \{ T_1, T_2, \ldots, T_n\}$ of the system, we can then
describe permutations over $\mathcal{T}$ as the composition of permutations
over the individual types, $\pi_{T_1} \circ \pi_{T_2} \circ \cdots \circ \pi_{T_n}$.
Let $\permset{\mathcal{T}}$ be the set of such ``system-wide'' permutations
over $\mathcal{T}$.

%% TODO: We do not elaborate on allowable types --- scalars, records, arrays etc
%% --- anywhere before in the paper. Their sudden introduction in the subsequent
%% paragraph is jarring.

ESM sketches and input and output channels may thus be parameterized by
symmetric values. The state variables and array variable indices\footnote{To
  apply a permutation $\pi$ to an array value, we first apply $\pi$ to each
  element of the array, and then permute the indices of the array itself.} of an
$\efsmsk$ may also be of symmetric type. Given the symmetric types $\mathcal{T}$
and an interpretation $R$ of the unknown functions in an $\efsmsk$ $A$, we say
that $A$ is \emph{symmetric} with respect to $\mathcal{T}$ if every execution
$(l_0, \sigma_0) \to (l_1, \sigma_1) \to\cdots\to (l_n, \sigma_n) \to\cdots$ of $A$
under $R$ also implies the existence of the permuted
execution $(\pi(l_0), \pi(\sigma_0)) \to (\pi(l_1), \pi(\sigma_0)) \to
\cdots (\pi(l_n), \pi(\sigma_n)) \to \cdots$ of $A$, where the channel
identifiers along transitions are also suitably permuted, for every
permutation $\pi \in \permset{\mathcal{T}}$.

We therefore require that any interpretation $R$ considered be such
that the completed $\efsmsk$ $A$ is symmetric with respect to
$\mathcal{T}$ under $R$. For every unknown function $f$ in $A$,
requiring that $\forall d \in \mathsf{dom}(f), \pi(f(d)) =
f(\pi(d)))$, for each permutation $\pi \in \permset{\mathcal{T}}$,
ensures that the behavior of $f$ is symmetric. To see why, let us
consider the example of the function $f_{\mathrm{turn}}$, which could
be used to update the variable $\mathrm{turn}$ in Peterson's
algorithm, shown in Figure~\subref*{figure:peterson_skeleton}. Suppose
that the variable $\mathrm{turn}$ was to be updated as $\mathrm{turn}
:= f_{\mathrm{turn}}(\mathtt{Pm}, \mathtt{Po}, \mathrm{flag},
\mathrm{turn})$. Since there are only two permutations possible in
this system ($\mathtt{Pm} = P_0, \mathtt{Po} = P_1$, and $\mathtt{Pm}
= P_1, \mathtt{Po} = P_0$), we would require, for example,
$f_{\mathrm{turn}}(P_1, P_0, P_0, \zug{\true,\false}) = P_0 \leftrightarrow
f_{\mathrm{turn}}(P_0, P_1, P_1, \zug{\false, true}) = P_1$. Observe,
that the two sides of the bi-directional implication can be obtained
from each other by the permutation $\pi \equiv \zug{P_0 \mapsto P_1,
P_1 \mapsto P_0}$. In general we would add such constraints for each
value in the range of the function, and for each permutation $\pi \in
\permset{\mathcal{T}}$. Section~\ref{sec:Completion}, provides some
additional examples to illustrate this.
Note that while we have
restricted the discussion here to only \emph{full
symmetry}, other notions of symmetry such as \emph{ring symmetry} and
\emph{virtual symmetry} can also be accommodated into our
formalization.

\subsection{Completion Problem}
In many cases, the designer has some prior knowledge about the unknown functions
used in an \efsmsk. For example, the designer may know that the variable
$\textit{turn}$ is read-only during the $(L_3, L_4)$ transition of
Peterson's algorithm. The designer may also know that the unknown guard of a
transition is independent of some state variable. Many instances of such ``prior
knowledge'' can already be expressed using the formalism just described: the
update expression of $\textit{turn}$ in the unknown transition can be set to the
identity function (in the first case), and the designer can omit the irrelevant
variable from the signature of the update function (in the second case).
We also allow the designer to specify additional constraints on the unknown
functions: she may know, as in the case of Peterson's algorithm for example, that
$g_{\mathrm{crit}}(\myid, \otherid, \textit{flag}, \textit{turn}) \lor
  g_{\mathrm{wait}}(\myid, \otherid, \textit{flag}, \textit{turn})$, for every
valuation of the function arguments $\myid$, $\otherid$, $\textit{flag}$, and
$\textit{turn}$. This additional knowledge, which is helpful to guide the
synthesizer, is encoded in the initial constraints $\Phi_0$ imposed on candidate
interpretations of $U$. Note that these constraints might refer to multiple
unknown functions from the same or different \efsmsk.

Formally, we can now state the completion problem as:
Given a set of \efsmsk $A_1, \ldots A_N$ with sets of unknown functions
$U_1, \ldots, U_N$, an environment \efsmsk $E$ with an empty set of unknown
functions, and a set of constraints $\Phi_0$ on the unknown functions $U =
U_1 \cup \cdots \cup U_N$, find an interpretation $\interpretation$ of $U$,
such that
\begin{inparaenum}[(\itshape a\upshape)]
  \item $A_1, \ldots, A_N$ are deterministic under $\interpretation$,
  \item the completed system $\Pi = A_1 \compose \cdots \compose A_N \compose E$
    is symmetric with respect to $\mathcal{T}$ under $\interpretation$, where
    $\mathcal{T}$ is the set of symmetric types in the system,
  \item $\interpretation$ satisfies the constraints in $\Phi_0$, and
  \item the product $\Pi$ under $\interpretation$ is deadlock-free, safe, and
    live.
\end{inparaenum}

\section{Solving the Completion Problem}
\label{sec:Completion}

%!TEX root = cav15.tex

% Completion algorithm
% 1. High-level strategy (+ block diagram of system)
% 2. Optimization problem
% 3. SyGuS post-processing
% 4. Optimizations

\begin{figure}[t]
\centering
\begin{tikzpicture}[node distance=1cm, every node/.style={draw, align=center, inner sep=.4cm}]
  \path
  node[basic box=blue] (initialize constraints) {Add input,\\determinism, and\\symmetry constraints}
  node[basic box=blue, below=7mmof initialize constraints] (solve constraints) {Solve constraints:\\Produce model for\\unknown functions}
  node[basic box=blue, right=5cm of solve constraints] (complete automata) {Instantiate protocol\\with interpretation}
  coordinate [at={($(initialize constraints)-(0,5mm)$)}] (temp point)
  coordinate [at={($(initialize constraints)+(0,5mm)$)}] (temp point2)
  node[input box, at=(complete automata |- temp point)] (partial automata) {\efsmsk\\ $A_1, A_2, \ldots, A_N$}
  node[input box, right=25mm of temp point2] (initial constraints) {Constraints $\Phi_0$\\ on unknown functions}

  node[decision=blue, at=($(solve constraints.east)!.4!(complete automata.west)$)] (solve constraints decision) {SAT?}
  node[basic box=blue, right=7mm of complete automata] (model check) {Model check\\protocol}
  node[input box, above=6mm of model check] (environment) {Environment\\\efsmsk $E$}
  node[decision=blue, below=7mm of model check] (model check decision) {Errors?}
  node[basic box=green, below=5mm of model check decision, inner sep=2mm] (success) {Correct\\Interpretation}
  node[basic box=blue, left=22mm of success] (sygus) {\sygus Solver}
  node[input box, left=15mm of sygus] (expressions) {Symbolic Expressions}
  %% node[basic box=blue, at={($(solve constraints)!.5!(model check decision)-(0,2.5cm)$)}] (analyze errors) {Analyze errors \&\\update constraints}
  node[basic box=blue, left=25mm of model check decision] (analyze errors) {Analyze errors \&\\update constraints}
  node[basic box=red, below=5mm of solve constraints decision, inner sep=2mm] (unsolvable) {No completion}
  ;
  \path[data line, ->]
  (initial constraints) edge ($(temp point2)+(18.5mm,0)$)
  (partial automata) edge ($(temp point)+(18.5mm,0)$)
  (partial automata -| complete automata) coordinate (temp point 2)
  (partial automata) edge (complete automata)
  (environment) edge (model check)
  (success) edge (sygus)
  (sygus) edge (expressions)
  ;

  \path[fat blue line,->]
  (model check decision) edge[fat blue line] node[above right=-3mm and -1mm, black, draw=none] {No?} (success)

  % (model check decision) edge[fat blue line, -] node[right, black, draw=none] {Yes?} (temp point)
  (model check decision) edge[fat blue line] node[above right=-2mm and -5mm, black, draw=none] {Yes?} (analyze errors)
  (analyze errors -| solve constraints) coordinate (temp point)
  (analyze errors) edge[fat blue line, -] (temp point)
  (temp point) edge[fat blue line] (solve constraints)
  (solve constraints) edge (solve constraints decision)
  (solve constraints decision) edge node[above right=-2mm and -17mm, black, draw=none] {Yes?\\Interpretation\\for unknown\\ functions}(complete automata)
  (solve constraints decision) edge node[below right=-6mm and -2mm , black, draw=none] {No?} (unsolvable)
  (complete automata) edge (model check)
  (model check) edge (model check decision)
  (initialize constraints) edge (solve constraints)
  ;
\end{tikzpicture}
\caption{Completion Algorithm.}
\label{figure:algorithm-flowchart}
\end{figure}
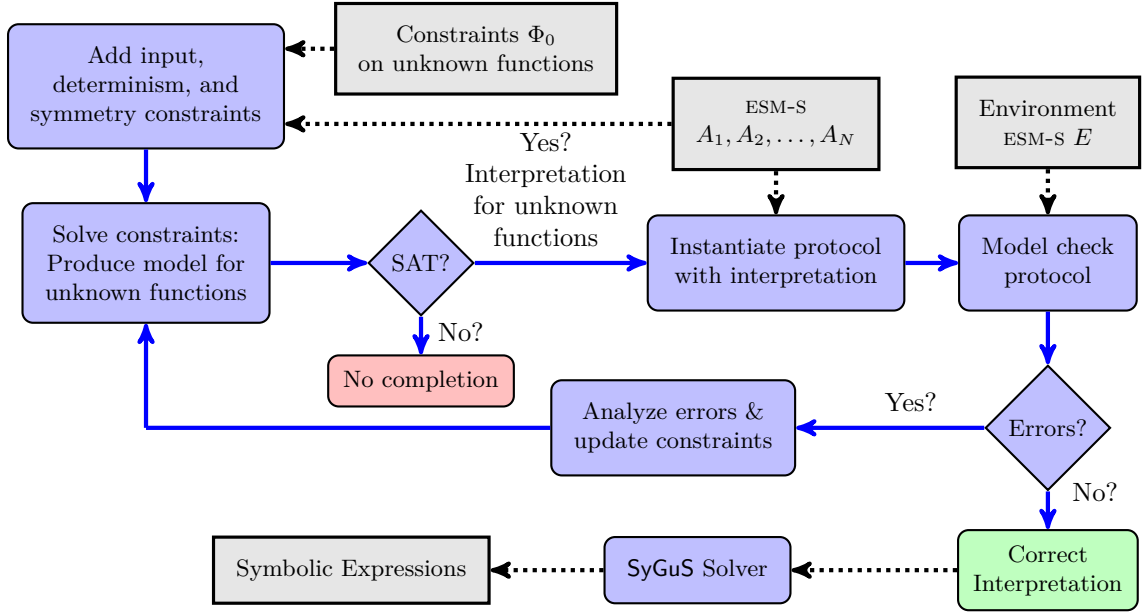

The synthesis algorithm is outlined in
Figure~\ref{figure:algorithm-flowchart}.  We maintain a set of
constraints $\Phi$ on possible completions, and repeatedly query an
SMT solver --- Z3~\cite{demoura-08} in our implementation --- for
candidate interpretations for the unknown functions satisfying all
constraints in $\Phi$. If the protocol instantiated with the candidate
interpretations is certified correct by the model checker, \ie, the
instantiated protocol satisfies all the safety and liveness
requirements set forth by the programmer, then we are done. Otherwise,
counter-example executions returned by the model checker are analyzed,
and $\Phi$ is strengthened with further constraints which eliminate
all interpretations which result in similar erroneous executions from
consideration in subsequent iterations of the algorithm.  If a
symbolic expression is required, we can submit the correct
interpretation to a \sygus solver~\cite{udupa-fmcad-sygus}. A \sygus
solver takes a set of constraints $\mathcal{C}$ on an unknown
functions $f$ together with the search space for the body of $f$ ---
expressed as a grammar --- and finds an expression in the grammar for
$f$, such that it satisfies the constraints $\mathcal{C}$.  In this
section, we first describe the initial determinism and symmetry
constraints expected of all completions. Next, we briefly describe the
model checker used in our implementation, and then describe how to
analyze counterexamples returned by the model checker.  Finally, we
describe additional heuristics to bias the SMT solver towards
intuitively simpler completions.

%% As discussed earlier, the algorithm we use to solve the completion is a CEGIS procedure.
%% Roughly this means that we maintain a set of constraints on possible completions and
%% follow an iterative approach.
%% In each iteration, we pick a solution that satisfies the set of constraints, test it
%% against the correctness properties, and in case it does not satisfy them,
%% we inform the set of constraints with the reason why it did not.
%% In this section, we will explain the different components of the scheme outlined above,
%% and highlight the parts of our solution that were instrumental to making it effective in the case of
%% the completion of EFSMS problem.

%% We start with by describing how we handle the different types of errors that might arise when
%% testing a candidate completion.
%% Those can be deadlocks, safety, or liveness violations.
\subsection{Initial Constraints}
\bfpara{Determinism Constraints.}{
Recall that an \efsmsk is deterministic under an interpretation $R$ if
and only if for every state $(l, \sigma)$ if there are multiple
transitions enabled at $(l, \sigma)$, then they must be input
transitions on distinct input channels. We constrain the
interpretations chosen at every step such that all ESM sketches in the
protocol are deterministic. Consider the \efsmsk for Peterson's
algorithm shown in Figure~\ref{figure:peterson_sketch}. We have two
transitions from the location $L_3$, with guards
$g_{\mathrm{crit}}(\texttt{Pm}, \texttt{Po}, \textit{flag}, \textit{turn})$ and
$g_{\mathrm{wait}}(\texttt{Pm}, \texttt{Po}, \textit{flag}, \textit{turn})$. We ensure
that these expressions never evaluate to true simultaneously with
the constraint $\neg\exists v_1 v_2 v_3 v_4
\left(g_{\mathrm{crit}}(v_1, v_2, v_3, v_4) \wedge
g_{\mathrm{wait}}(v_1, v_2, v_3, v_4)\right)$. Although this is a
quantified expression, which can be difficult for SMT solvers to
solve, note that we only support finite types, whose domains are often
quite small. So our tool unrolls the quantifiers and presents only
quantifier-free formulas to the SMT solver.
}

\bfpara{Symmetry Constraints.}{
Suppose that the interpretation chosen for the guard $g_{\mathrm{crit}}$
shown in Figure~\ref{figure:peterson_sketch}, was such that
$g_{\mathrm{crit}}(\texttt{P0}, \texttt{P1}, \langle\bot, \top\rangle,
\texttt{P0}) = \texttt{true}$. Then for the ESM sketch to be symmetric
under this interpretation, we require that $g_{\mathrm{crit}}(\texttt{P1},
\texttt{P0}, \langle\top, \bot\rangle, \texttt{P1} = \texttt{true}$ as
well, because the latter expression is obtained by applying the permutation
$\{\texttt{P0} \mapsto \texttt{P1}, \texttt{P1} \mapsto \texttt{P0}\}$ on
the former expression. Note that the elements of the \emph{flag} array in
the preceding example were flipped, because \emph{flag} is an array indexed
by the symmetric type $\mathsf{processid}$. In general, given a function $f
\in U_i$, we enforce the constraint $\forall \pi \in \permset{\mathcal{T}}
\forall d \in \mathsf{dom}(f) (f(\pi(d)) \equiv \pi(f(d)))$, where
$\mathcal{T}$ is the set of symmetric types that appear in $A_i$. As in the
case of determinism constraints, we unroll the quantifiers here as well.
}

\subsection{Model Checker}
To effectively and repeatedly generate constraints to drive
the synthesis loop, a model checker needs to:
(a) support checking liveness properties, with
algorithmic support for fine grained notions of strong and weak
fairness, (b) dynamically prioritize certain paths over
others (\emph{cf.}
Section~\ref{subsec:optimizations_and_heuristics}), and (c)
exploit symmetries inherent in the model. The fine grained notions of
fairness over sets of transitions, rather than bulk process fairness
are crucial. For instance, in the case of unordered channel processes,
we often require that no message be delayed indefinitely, which cannot
be captured by enforcing fairness at the level of the entire
process. The ability to prioritize certain paths over others is also
crucial so that candidate interpretations are exercised to the extent
possible in one model checking run (\emph{cf.}
Section~\ref{subsec:optimizations_and_heuristics}). Finally, support
for symmetry-based state space reductions, while not absolutely
crucial, can greatly speed up each model checking run.

Surprisingly, we found that none of the well-supported model checkers
met all of our requirements. \textsc{Spin}~\cite{holzmann-97} only
supports weak process fairness at an algorithmic level and does not
employ symmetry-based reductions. Our efforts to encode the necessary
fine grained strong fairness requirements as \textsc{ltl} formulas in
\textsc{Spin} resulted in the B\"uchi monitor construction step either
blowing up or generating extremely large monitor processes. Support
for symmetry-based reductions is present in
Mur$\varphi$~\cite{dill-96a,dill-96b}, but it lacks support for
liveness checking.~\footnote{There exists an unmaintained version of
Mur$\varphi$ which does support checking of some restricted forms of
\textsc{ltl} properties, but it only supports weak fairness.}
SMC~\cite{sistla-00} is a model checker with support for symmetry
reduction and strong and weak process fairness. Unfortunately, it is
no longer maintained, and has very rudimentary counterexample
generation capabilities. Finally, \textsc{NuSMV}~\cite{cimatti-02}
does not support symmetry reductions, but supports strong and weak
process level fairness. But, due to bugs in the implementation
of counterexample generation, we were unable to obtain counterexamples
in some cases.

We therefore implemented a model checker based on the ideas
used in Mur$\varphi$~\cite{dill-96b} for symmetry reduction, and an
adaptation of the techniques presented in~\cite{sistla-97} for
checking liveness properties under fine grained fairness
assumptions. We plan on releasing the model checker as standalone
open-source tool in the near future.
The model checking algorithm consists of the following steps:
\begin{enumerate}[itemsep=1pt,topsep=2pt,parsep=2pt]
\item
Construct the symmetry reduced state graph in the form of the
\emph{Annotated Quotient Structure} described in earlier
literature~\cite{sistla-97}. In
our setting, we treat the liveness monitors the same way as other
processes in the system. So this graph is really the symmetry
reduced \emph{product structure} described in earlier
literature~\cite{sistla-97}. The task is to find \emph{fair},
\emph{accepting} cycles in this symmetry reduced product graph.
\item
To accomplish this search for fair cycles in a sound and complete way,
we implicitly search over the \emph{threaded} graph described in
earlier literature~\cite{sistla-97}. This threaded graph annotates
each state with an additional component which keeps track of the
permutations that have been applied along a path, which will then be
used to ``adjust'' the fairness requirements that have already been
satisfied. This is necessary, because the product state space is
\emph{compressed}, as a result, a path in the product state space
could correspond to more than one uncompressed paths. We refer the
reader to published literature~\cite{sistla-97} for more details on
the construction of the \emph{threaded} graph structure.
\item
Once the \emph{threaded} graph has been constructed, we then simply
compute the accepting strongly connected components (SCCs) in this
threaded graph using Tarjan's algorithm for computing strongly
connected components in a directed graph~\cite{tarjan-72}. An
\emph{accepting} SCC is simply an SCC which contains at least one
accepting state.
\item
In each accepting SCC $C$, where all the weak fairness requirements
are satisfied, and for each strong fairness requirement $F_s$ which is
not satisfied in $C$, \ie, some transition in $F_s$ is \emph{enabled}
in $C$, but no transition in $F_s$ is ever \emph{taken} in $C$, we delete
from the threaded graph, all the states where some transition in $F_s$
is enabled.
\item
Steps (3) and (4) are repeated until either (a) an accepting SCC $C$
is found which satisfies all the strong and weak fairness
requirements, in which case we can construct a counterexample from
$C$, or (b) no more accepting SCCs remain, in which case the protocol
satisfies all the liveness requirements and no \emph{fair},
\emph{accepting} execution exists.
\end{enumerate}

\subsection{Analysis of Counterexamples}
We now describe our algorithms for analyzing counterexamples by way of
examples first and then provide a formal description of the algorithms.

\bfpara{Analyzing deadlocks, an example.}{
In Figure~\ref{figure:peterson_sketch}, consider the candidate
interpretation where both $g_\textrm{crit}$, $g_\textrm{wait}$ are set
to be universally $\false$.  Two deadlock states are then
reachable:
$S_1 = ((L_3, L_3), \{\textit{flag}\mapsto \langle \top, \top\rangle,
\textit{turn}\mapsto \texttt{P1}\}$
and
$S_2 = ((L_3, L_3), \{\textit{flag}\mapsto \langle \top, \top\rangle,
\textit{turn}\mapsto \texttt{P0}\}$.
We strengthen $\Phi$ by asserting that these deadlocks do not occur in
future interpretations: either $S_1$ is unreachable, or the system can
make a transition from $S_1$ (and similarly for $S_2$).  In this
example, the reachability of both deadlock states is not dependent on
the interpretation, \emph{i.e.}, the execution that leads to the states does
not exercise any unknown function, hence, we need to make sure that
the states are not deadlocks.  The possible transitions out of
location $(L_3, L_3)$ are the transitions from $L_3$ to $L_3$ (waiting
transition) and from $L_3$ to $L_4$ (critical transition) for each of
the two processes.  In each deadlock state, at least one of the four
guards has to be true:
$g_\textrm{wait}(\texttt{P0}, \texttt{P1}, \langle\top, \top\rangle,
\texttt{P1}) \lor g_\textrm{crit}(\texttt{P0}, \texttt{P1},
\langle\top, \top\rangle, \texttt{P1}) \lor
g_\textrm{wait}(\texttt{P1}, \texttt{P0}, \langle\top, \top\rangle,
\texttt{P1}) \lor g_\textrm{crit}(\texttt{P1}, \texttt{P0},
\langle\top, \top\rangle, \texttt{P1})$
for $S_1$, and
$g_\textrm{wait}(\texttt{P0}, \texttt{P1}, \langle\top, \top\rangle,
\texttt{P0}) \lor g_\textrm{crit}(\texttt{P0}, \texttt{P1},
\langle\top, \top\rangle,$ $\texttt{P0}) \lor
g_\textrm{wait}(\texttt{P1}, \texttt{P0}, \langle\top, \top\rangle,
\texttt{P0}) \lor g_\textrm{crit}(\texttt{P1}, \texttt{P0},
\langle\top, \top\rangle, \texttt{P0})$
for $S_2$.  The two disjunctions are added to the set of constraints,
since any candidate interpretation has to satisfy them in order for
the resulting product to be deadlock-free.
}

\bfpara{Analyzing safety violations, an example.}{
Consider now an erroneous interpretation where the critical transition
guards are \true for both processes when $\textit{turn}$ is
$\texttt{P0}$, that is:
$g_\textrm{crit}(\texttt{P0}, \texttt{P1}, \langle \top, \top \rangle,
\texttt{P0})$
and
$g_\textrm{crit}(\texttt{P1}, \texttt{P0}, \langle \top, \top \rangle,
\texttt{P0})$
are set to \true.  Under this interpretation the product can reach the
error location $(L_4, L_4)$. We perform a weakest precondition
analysis on the corresponding execution to obtain a necessary
condition under which the safety violation is possible. In this case,
the execution crosses both critical transitions and the generated
constraint is
$\lnot g_\textrm{crit}(\texttt{P0}, \texttt{P1}, \langle \top, \top
\rangle, \texttt{P0}) \lor \lnot g_\textrm{crit}(\texttt{P1},
\texttt{P0}, \langle \top, \top \rangle, \texttt{P0})$.
Note that the conditions obtained from this analysis are necessary;
the product under any interpretation that does not satisfy them will
exhibit the same safety violation.
}

\bfpara{Analyzing liveness violations, an example.} {
An interpretation that satisfies the constraints gathered above is one
that, when \emph{turn} is $\texttt{P0}$, enables both waiting transitions
and disables the critical ones.  Intuitively, under this
interpretation, the two processes will not make progress if \emph{turn}
is \texttt{P0} when they reach $L_3$.  The executions in which the
processes are at $L_3$ and either $P_0$ or $P_1$ continuously take the
waiting transition is an accepting one.  As with safety violations, we
eliminate liveness violations by adding constraints generated through
weakest precondition analysis of the accepting executions.  In this
case, this results in two constraints:
$\lnot g_\mathrm{wait}(\texttt{P0}, \texttt{P1}, \langle\top,
\top\rangle, \texttt{P0})$
and
$\lnot g_\mathrm{wait}(\texttt{P1}, \texttt{P0}, \langle\top,
\top\rangle, \texttt{P0})$.
However, in the presence of fairness assumptions, these constraints
are too strong.  This is because removing an execution that causes a
fair liveness violation is not the only way to resolve it: another way
is to make it unfair.  Given the weak fairness assumption on the
transitions on the $\text{critical}_\texttt{Pi}$ channels, the correct
constraint generated for the liveness violation of Process $P_0$ is:
$\lnot g_\mathrm{wait}(\texttt{P0}, \texttt{P1}, \langle \top, \top
\rangle, \texttt{P0}) \lor g_\mathrm{crit}(\texttt{P0}, \texttt{P1},
\langle \top, \top \rangle, \texttt{P0}) \lor
g_\mathrm{crit}(\texttt{P1}, \texttt{P0}, \true, \true, \texttt{P0})$,
where the last two disjuncts render the accepting execution unfair.
}

We now describe in detail how we perform the analysis of counterexamples
returned by the model checker.  Our implementation first composes the ESM
sketches to form a \emph{product} \efsmsk $\Pi$. It then compiles down this
product \efsmsk $\Pi$ into guarded commands, where the guards and updates
are as defined in Section~\ref{subsection:efsms}. The
guards and updates of the guarded commands are also transformed by the
compiler to use $\mathsf{select}$, $\mathsf{store}$, $\mathsf{project}$ and
record $\mathsf{update}$ functions\footnote{These are functions defined in the
  theory of arrays and records by the SMTLIB2 standard. For details, see
  \url{http://smt-lib.org/}} for reads and updates of arrays and records
respectively.  Furthermore, repeated assignments to the same variable in a
guarded command are coalesced into a single assignment. In effect, each
variable (be it of a scalar type, an array type or a record type) have
\emph{only one} assignment to it in the list of updates associated with
each guarded command.  These transformations on the guarded commands make
it easier to compute the weakest preconditions of predicates with respect
to the guarded commands, as we shall now explain.

Let the set of guarded commands be $G$, given a guarded command $\cmd
\in G$, we define $\guardcmd{\cmd}$ to be the guard of $\cmd$ and
$\updatecmd{\cmd}$ to be the list of coalesced updates of $\cmd$. The
weakest precondition of a predicate $\varphi$ with respect to an
assignment statement $stmt \triangleq l := e$ is defined as
$\weakestpre{stmt}{\varphi} \equiv \varphi[l \gets e]$, where $\varphi[l
  \gets e]$ is the expression obtained by replacing all instances of the
sub-expression $l$ in $\varphi$ with the expression $e$. We extend the
definition of the weakest precondition of a predicate $\varphi$
with respect to a sequence of statements in the natural
way. The weakest precondition of a predicate $\varphi$ with respect
to a guarded command $\cmd$ is defined is defined as
$\weakestprecmd{\cmd}{\varphi} \equiv \guardcmd{cmd} \rightarrow
\weakestpre{\updatecmd{\cmd}}{\varphi}$.

\bfpara{Analyzing Safety and Deadlock Counterexamples.}{
  Given an error trace (\emph{i.e.}, a witness for a safety violation
or a deadlock) which consists of an initial state valuation
$\sigma_0$, and a sequence of guarded commands from $G$, say, $\cmd_1,
\cmd_2, \ldots, \cmd_{\text{n}}$. We define $\mathsf{pre}_0(\varphi)
\equiv \varphi$, and recursively define
$\mathsf{pre}_{\text{i}}(\varphi) \equiv \weakestprecmd{\cmd_{\text{n} -
\text{i} - 1}}{\mathsf{pre}_{\text{i-1}}(\varphi)}$. Then, if the trace
is a witness for a safety violation, we add the constraint
$\mathsf{pre}_{\text{n}}(\false)[v \gets \sigma_0(v)]$, for every $v \in
V_1 \cup V_2 \cup \ldots \cup V_n$, to our set of constraints $\Phi$,
which essentially ensures that the particular execution is no longer
possible under all interpretations chosen for the set of unknown
functions $U$ in the future. On the other hand if the trace is a
witness for a deadlock, we add
$\mathsf{pre}_{\text{n}}\left(\bigvee_{\cmd \in
G}\guardcmd{\cmd}\right)[v \gets \sigma_0(v)]$, for every $v \in
V_1 \cup V_2 \cup \ldots \cup V_n$, to $\Phi$. This constraint
ensures that if
this particular execution is ever permitted under an interpretation
for the unknown functions $U$ chosen in the future, then some guarded
command is enabled at the end of the execution, under that
interpretation, therefore no longer rendering the final state of the
execution a deadlock.
}

\bfpara{Analyzing Liveness Counterexamples.}{
  We assume that infinite accepting
  executions are given as a pair of a finite stem execution of size
  $n$ and a finite cycle execution of size $m$.  First, we describe
  the case where no fairness assumptions exist in the system.  The
  constraint computed from an accepting execution asserts either that
  the sequence of transitions should not be enabled or that the state
  of the system at the beginning of the cycle should be not be the
  same as the state at the end.  If the set of variables of $\Pi$ is
  $\{s_1, \ldots, s_N\}$ we introduce symbolic constants
  $s_1', \ldots, s_N'$ and set
  $\phi \equiv s_1 \neq s_1' \lor s_2 \neq s_2' \lor \cdots \lor s_N
  \neq s_N'$.
  We first compute $\phi' = \mathsf{pre}_{\text{m}}(\phi)$ on the
  cycle execution and then substitute $s_1', \ldots, s_N'$ for
  $s_1, \ldots, s_N$ in $\phi'$:
  $\phi'' = \phi'[s_1' \gets s_1, \ldots, s_N' \gets s_N]$.  We then
  get the final constraint by computing
  $\mathsf{pre}_{\text{n}}(\phi'')$ on the stem execution.

  We now describe the case where strong fairness assumptions are
  present.  The treatment of weak fairness assumptions is similar.
  Let $\mathcal{F}$ be the set of strong fairness assumptions and $G$
  be the union of all fairness sets $F\in\mathcal{F}$ such that every
  guarded command in $F$ is disabled in the cycle.  We adapt the
  computation of $\mathsf{pre}_\text{i}$ in the cycle execution as
  follows:
  $\mathsf{pre'}_{\text{i}}(\varphi) \equiv \weakestprecmd{\cmd_{\text{n}
      - \text{i} - 1}}{\mathsf{pre}_{\text{i-1}}(\varphi) \lor
    \bigvee_{\cmd \in F}\guardcmd{\cmd}}$.
  Enabling a command $\cmd$ in $G$ at a step in the cycle execution
  has the effect of making the accepting cycle unfair: since $\cmd$ is
  never executed in the cycle, enforcing $\guardcmd{\cmd}$ makes
  $\cmd$ infinite often enabled but never taken.
}

\subsection{Optimizations and Heuristics.}
\label{subsec:optimizations_and_heuristics}
We describe a few key optimizations and heuristics that improve
the scalability and predictability of our technique.

\bfpara{Not all counterexamples are created equal.}{
  The constraint we get from a single counter-example trace is weaker when it
  exercises a large number of unknown functions. Consider, for example, a
  candidate interpretation for the incomplete Peterson's algorithm which, when
  $\textit{turn} = \texttt{P0}$, sets both waiting transition guards
  $g_\mathrm{wait}$ to $\true$, and both critical transition guards
  $g_\mathrm{crit}$ to $\false$. We have already seen that the product is not
  live under this interpretation. From the infinite execution leading up-to the
  location $(L_3, L_3)$, and after which $P_0$ loops in $L_3$, we obtain the
  constraint\footnote{Ignoring fairness assumptions.} $\lnot g_\mathrm{wait}
    (\texttt{P0}, \texttt{P1}, \langle \top, \top \rangle, \texttt{P0})$.
  On the other hand, if we had
  considered the longer self-loop at $(L_3, L_3)$, where $P_0$ and $P_1$
  alternate in making waiting transitions, we would have obtained the
  weaker constraint $\lnot g_\mathrm{wait}(\texttt{P0}, \texttt{P1},
    \langle \top, \top \rangle, \texttt{P0}) \lor \lnot g_\mathrm{wait}
    (\texttt{P1}, \texttt{P0}, \langle \top, \top \rangle, \texttt{P0})$.
  In general, erroneous traces which exercise fewer unknown functions
  have the potential to prune away a larger fraction of the search
  space and are therefore preferable over traces exercising a larger
  number of unknown functions.

  In each iteration, the model checker discovers several erroneous
states. In the event that the candidate interpretation chosen is
blatantly incorrect, it is infeasible to analyze paths to all error
states. A na\"ive solution would be to analyze paths to the first $n$
errors states discovered (where $n$ is configurable). But depending on
the strategy used to explore the state space, a large fraction these
errors could be similar\footnote{We observed this phenomenon in our
initial experiments.}, and would only provide us with rather weak
constraints. On the other hand, exercising as many unknown functions as
possible, along different paths, has the potential to provide stronger
constraints on future interpretations.  In summary, we bias the model
checker to \emph{cover} as many unknown functions as possible, such
that each path exercises as few unknown functions as possible.
}

\bfpara{Heuristics/Prioritizations to guide the SMT solver.}{
As mentioned earlier, we use an SMT solver to obtain interpretations
for unknown functions, given a set of constraints. When this set is
small, as is the case at the beginning of the algorithm, there exist
many satisfying interpretations. At this point the interpretation
chosen by the SMT solver can either lead the rest of the search down a
``good'' path, or lead it down a futile path. Therefore the run time of the
synthesis algorithm can depend heavily on the interpretations returned
by the SMT solver, which we consider a non-deterministic black box in
our approach.

To reduce the influence of non-determinism of the SMT solver on the
run time of our algorithm, we bias the solver towards
specific forms of interpretations by asserting additional
constraints. These constraints associate a
\emph{cost} with interpretations and require an interpretation with a
given bound on the cost, which is relaxed whenever the SMT solver
fails to find a solution.

We briefly describe the most important of the heuristics/prioritization
techniques: (1) We minimize the number of points in the domain of an
unknown guard function at which it evaluates to true. This results in
minimally permissive guards. (2) Based on the observation that
most variables are unchanged in a given transition, we
prioritize interpretations where update functions
leave the value of the variable unchanged, as far as possible.
(3) In the event that the value of the variable cannot be left
unchanged, we try to minimize the number of arguments on which the
value of the unknown function depends, in an attempt to bias the SMT
solver towards \emph{intuitively} simple interpretations.
}

\section{Experiments}
\label{sec:experiments}
\subsection{Peterson's Mutual Exclusion Protocol}

% TODO try removing more things from the protocol?

In addition to the missing guards $g_\mathrm{grit}$ and
$g_\mathrm{wait}$, we also replace the update expressions of
$\text{flag}[\myid]$ in the $(L_1, L_2)$ and $(L_4, L_1)$ transitions
with unknown functions that depend on all state variables.
In the initial constraints we require that
$g_{\mathrm{crit}}(\myid, \otherid, \textit{flag}, \textit{turn}) \lor
g_{\mathrm{wait}}(\myid, \otherid, \textit{flag}, \textit{turn})$.
The synthesis algorithm returns with an interpretation in less than a
second.  Upon submitting the interpretation to a \sygus solver, the
synthesized expressions match the ones shown in
Figure~\ref{figure:peterson_sketch}.

\subsection{Self-stabilizing Systems}
\label{subsec:self_stabilizing_systems}
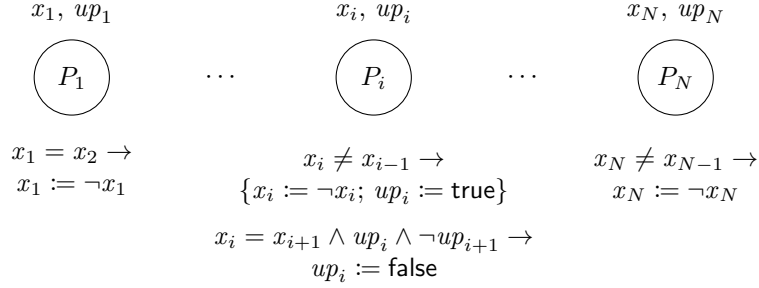
\begin{figure}[t!]
\centering
\begin{tikzpicture}
  \path
  node[state, minimum size=1cm] (P1) {$P_1$}
  node[state, minimum size=1cm, right=3cm of P1] (Pi) {$P_i$}
  node[state, minimum size=1cm, right=3cm of Pi] (PN) {$P_N$}

  node[at={($(P1)!.5!(Pi)$)}] (dots) {$\cdots$}
  node[at={($(Pi)!.5!(PN)$)}] (dots) {$\cdots$}

  % state
  node[above=.1cm of P1] {$\x_1$, $\up_1$}
  node[above=.1cm of Pi] {$\x_i$, $\up_i$}
  node[above=.1cm of PN] {$\x_N$, $\up_N$}

  % rules
  node[align=center, below=.3cm of PN] {$\x_N \neq \x_{N-1} \to$\\$\x_N \coloneqq \lnot \x_N$}
  node[align=center, below=.3cm of P1] {$\x_1 = \x_2 \to$\\$\x_1 \coloneqq \lnot \x_1$}
  node[align=center, below=.3cm of Pi] {
    $\x_i \neq \x_{i-1} \to$\\
    \{$\x_i \coloneqq \lnot \x_i$;
    $\up_i \coloneqq \true$\}\\[.2cm]

    $\x_i = \x_{i+1} \land \up_i \land \lnot \up_{i + 1} \to$\\
    $\up_i \coloneqq \false$\\[.2cm]
  }
  ;
\end{tikzpicture}
\caption{Self-stabilizing system processes.}
\label{fig-dijkstra}
\end{figure}

Our next case study is the synthesis of self-stabilizing
systems~\cite{dijkstra}. A distributed system is self-stabilizing if,
starting from an arbitrary initial state, in each execution, the
system eventually reaches a global \emph{legitimate} state, and only
legitimate states are ever visited after. We also require that every
legitimate state be reachable from every other legitimate
state. Consider $N$ processes connected in a line.  Each process
maintains two Boolean state variables $\x$ and $\up$.  The processes
are described using guarded commands of the form, ``$\texttt{if }
\textit{guard} \texttt{ then } \textit{update}$''.  Whether a command
is enabled is a function of the variable values $\x$ and $\up$ of the
process itself, and those of its neighbors. We attempted to synthesize
the guards and updates for the middle two processes of a four process
system ${P_1, P_2, P_3, P_4}$.  Specifically, the \efsmsk for $P_2$
and $P_3$ have two transitions, each with an unknown function as a
guard and two unknown functions for updating its state variables.  The
guard is a function of $\x_{i - 1}$, $\x_i$, $\x_{i + 1}$, $\up_{i -
1}$, $\up_{i}$, $\up_{i + 1}$, and the updates of $\x_i$ and $\up_i$
are functions of $\x_i$ and $\up_i$.  We followed the definition
in~\cite{tiwari} and defined a state as being legitimate if exactly
one guarded command is enabled globally. We also constrain the
completions of $P_2$ and $P_3$ to be identical. The complete self-stabilizing system is
shown in Figure~\ref{fig-dijkstra}.  In our experiment we synthesized
the guards and updates of processes $P_2$ and $P_3$ in a four process
system, i.e., $N=4$.

% In our experiments, we fix the number of processes to four $\{P_1, P_2, P_3, P_4\}$.
% We constructed an EFSMS for each of the middle processes, $P_2$ and $P_3$.
% The EFSMS has two transitions, each with an unknown functions as a guard and
% two unknown functions for updating its state variables.
% We used the same unknown functions in both EFSMS in order to
% produce a symmetric completion.
% The signature of the guards is $\boolean{}^6 \mapsto \boolean$, where
% the arguments are the state variables of
% the process and its neighbors:
% $\x_{i - 1}$, $\x_i$, $\x_{i + 1}$,
% $\up_{i - 1}$, $\up_{i}$, $\up_{i + 1}$, and
% the signature of the updates is $\boolean^2 \mapsto \boolean$ with arguments
% the current state values: $\x_i$, $\up_i$.
% We expressed the correctness properties described in the previous paragraph
% using liveness and safety EFSMS, and using additional constraints $\Phi$
% on the unknown functions.

Due to the large number of unknown functions needed to be synthesized
in this experiment and, in particular, because there were a lot of
input domain points at which the guards had to be true, the heuristic
that prefers minimally permissive guards, described in
Section~\ref{sec:Completion}, was not effective. However, the
heuristic that prioritizes interpretations in which the guards depend
on fewer arguments of their domain was effective. For state variable
updates, we applied the heuristic that prioritizes functions that
leave the state unchanged or set it to a constant. After passing the
synthesized interpretation through a \sygus solver, the expressions we
got were exactly the same as the ones found in~\cite{dijkstra}, and
presented in Figure~\ref{fig-dijkstra}.

\subsection{Cache Coherence Protocol}
\label{subsec:experiments_cache_coherence}
\begin{figure}[!t]
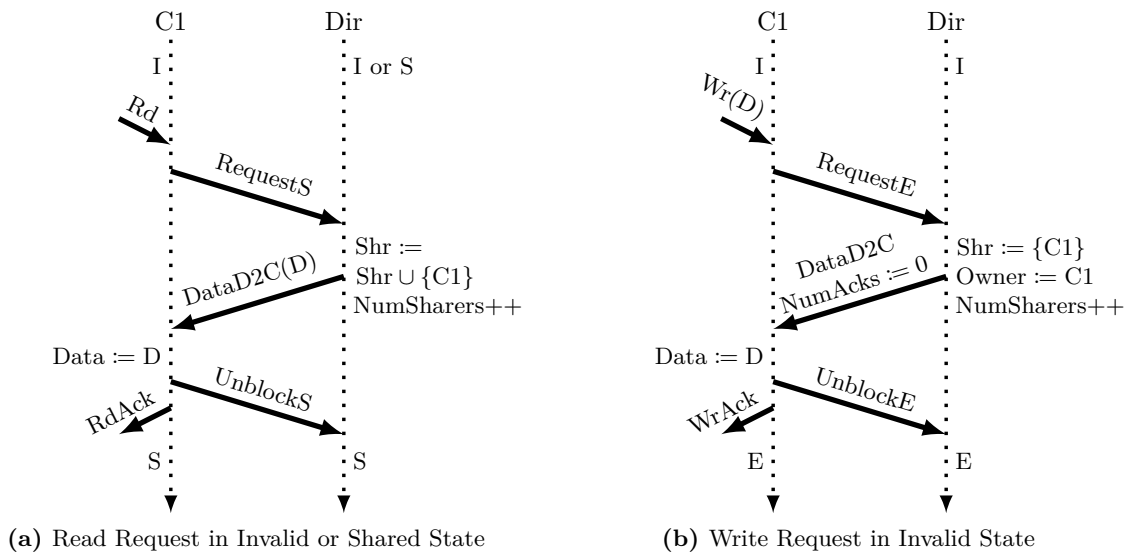

\centering
\subfloat[Read Request in Invalid or Shared State]{
  \input{figures/msi/msi_scenario_header.tex}

\begin{tikzpicture}
  \begin{scope}[x=2.3cm, y=-.7cm]
    \path
	(0, 0) node[above] {C1}
	(1, 0) node[above] {Dir}
  	(0, .5) node[left, lane label] {I}
	(1, .5) node[right, lane label] {I or S}
	(0, 0) edge[lane] (0, 9)
	(1, 0) edge[lane] (1, 9)
	(-.3, 1.5)
	edge[message]
	node[message node,pos=0.3] {Rd}
	(0, 2)

	(0, 2.5)
	edge[message]
	node[message node] {RequestS}
	(1, 3.5)

    (1, 4.5) node[right, command node]
    {
    \hspace{-13mm}$\text{Shr}\coloneqq$\\
    \hspace{-6mm}$\text{Shr}\cup \{ \text{C1} \}$\\
    $\text{NumSharers}\mathtt{++}$
    }

	(1, 4.5)
	edge[message]
 	node[message node] {DataD2C(D)}
	(0, 5.5)

	(0, 6) node[left, command node] {$\text{Data}\coloneqq \text{D}$}

	(0, 6.5)
    edge[message]
    node[message node] {UnblockS}
    (1, 7.5)

    (0, 7)
    edge[message]
    node[message node,pos=0.8] {RdAck}
    (-.3, 7.5)

    (1, 8) node[right, lane label] {S}
    (0, 8) node[left, lane label] {S}
    ;
  \end{scope}
\end{tikzpicture}
  \label{figure:msi_load_simple}
}
\subfloat[Write Request in Invalid State]{
  \input{figures/msi/msi_scenario_header.tex}
\begin{tikzpicture}
  \begin{scope}[x=2.3cm, y=-.7cm]
    \path
	(0, 0) node[above] {C1}
	(1, 0) node[above] {Dir}
  	(0, .5) node[left, lane label] {I}
	(1, .5) node[right, lane label] {I}
	(0, 0) edge[lane] (0, 9)
	(1, 0) edge[lane] (1, 9)

	(-.3, 1.5)
	edge[message]
	node[message node,pos=0] {Wr(D)}
	(0, 2)

	(0, 2.5)
	edge[message]
	node[message node] {RequestE}
	(1, 3.5)

    (1, 4.5) node[right, command node]
    {
    \hspace{-5mm}$\text{Shr}\coloneqq\{ \text{C1}\}$\\
    \hspace{-4mm}$\text{Owner} \coloneqq \text{C1}$\\
    $\text{NumSharers}\mathtt{++}$
    }

	(1, 4.5)
	edge[message]
 	node[message node]
 	{DataD2C\\$\text{NumAcks}\coloneqq 0$}
	(0, 5.5)
	(0, 6) node[left, command node] {$\text{Data}\coloneqq \text{D}$}

	(0, 6.5)
    edge[message]
    node[message node] {UnblockE}
    (1, 7.5)

    (0, 7)
    edge[message]
    node[message node,pos=0.8] {WrAck}
    (-0.3, 7.5)

    (1, 8) node[right, lane label] {E}
    (0, 8) node[left, lane label] {E}
    ;
  \end{scope}
\end{tikzpicture}
  \label{figure:msi_store_simple}
}
\caption{Simple Cases for Read and Write Requests}
\label{figure:msi_scenarios_1}
\end{figure}

\begin{figure}[!t]
\centering
\input{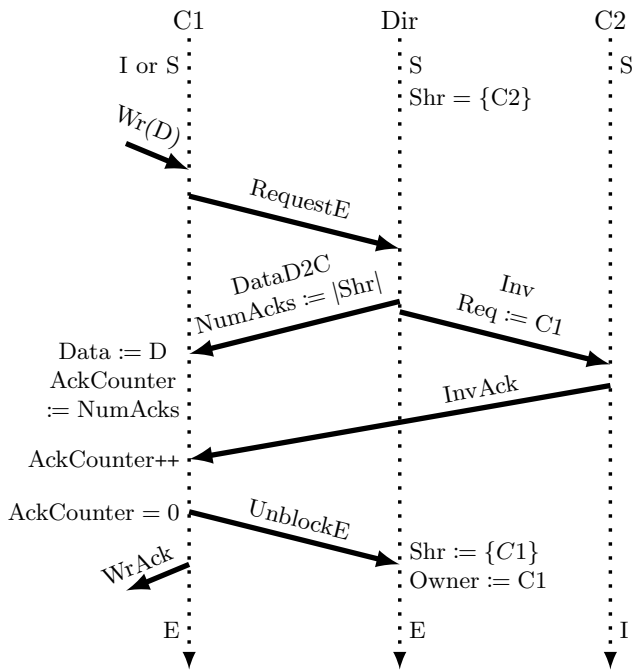}
\begin{tikzpicture}
  \begin{scope}[x=2.8cm, y=-.7cm]
    \path
	(0, 0) node[above] {C1}
	(1, 0) node[above] {Dir}
	(2, 0) node[above] {C2}
	(0, 0) edge[lane] (0, 12)
	(1, 0) edge[lane] (1, 12)
	(2, 0) edge[lane] (2, 12)

  	(0, .5) node[left, lane label] {I or S}
	(1, .5) node[right, lane label] {S}
	(2, .5) node[right, lane label] {S}

    (1, 1.15) node[right, lane label] {$\text{Shr} = \{\text{C2}\}$}
	(-.3, 2)
	edge[message]
	node[message node,pos=0.2] {Wr(D)}
	(0, 2.5)

	(0, 3)
	edge[message]
	node[message node] {RequestE}
	(1, 4)

	(1, 5)
	edge[message]
 	node[message node]
    {DataD2C\\$\text{NumAcks}\coloneqq |\text{Shr}|$}
	(0, 6)
	(0, 6.5) node[left, command node]
    {
      $\text{Data} \coloneqq \text{D}$\\
      $\text{AckCounter}$\\
      $\coloneqq\text{NumAcks}$
    }

	(1, 5.2)
	edge[message]
 	node[message node]
 	{Inv\\$\text{Req}\coloneqq \text{C1}$}
	(2, 6.2)

    (2, 6.6)
    edge[message]
    node[message node, pos=.3] {InvAck}
    (0, 8)

    (0, 8) node[left, command node]
    {$\text{AckCounter}$\texttt{++}}

    (0, 9) node[left, lane label] {$\text{AckCounter} = 0$}

    (0, 9)
    edge[message]
    node[message node] {UnblockE}
    (1, 10)

    (0,10)
    edge[message]
    node[message node,pos=0.7] {WrAck}
    (-.3, 10.5)

    (1, 10)
    node[right, command node]
    {
      \hspace{-1mm}$\text{Shr} \coloneqq \{C1\}$\\
      $\text{Owner} \coloneqq \text{C1}$
    }

    (0, 11.25) node[left, lane label] {E}
    (1, 11.25) node[right, lane label] {E}
	(2, 11.25) node[right, lane label] {I}
    ;
  \end{scope}
\end{tikzpicture}
\caption{Write Request in Shared State}
\label{figure:msi_store_inv}
\end{figure}

A cache coherence protocol ensures that the copies of shared data in
the private caches of a multiprocessor system are kept up-to-date with
the most recent version. We describe the working of a variant of the
German cache coherence protocol, which is often used as a case study
in model checking research~\cite{chou-04,talupur-08}. The protocol
consists of a \emph{Directory} process, $n$ symmetric \emph{Cache}
processes and $n$ symmetric \emph{Environment} processes, one for each
cache process. Each cache may be in the E, S or I state, indicating
read-write, read, and no permissions on the data respectively. All
communication between the caches and the directory is non-blocking,
and occurs over buffered, unordered communication channels.

Figure~\subref*{figure:msi_load_simple} shows the actions performed by the
various processes when a cache receives a \emph{read} command from its
environment. It sends a \emph{RequestS} message to the directory. In
this particular scenario, the directory has recorded that all other
caches are either in the I or S state, and proceeds to send the most
up-to-date copy of the data in a \emph{DataD2C} message. The cache
then updates its local copy of the data, notifies its environment that
the request has been satisfied and transitions to the S state.

Figure~\subref*{figure:msi_store_simple} shows what happens when a cache
receives a \emph{write} command from its environment along with the
new data value D to write. In this particular case, the directory
knows that all other caches are in the I state and thus proceeds to
acknowledge the \emph{RequestE} message from the cache with a
\emph{DataD2C} message which also contains the number of
acknowledgments the cache needs to wait for before gaining write
permissions on the data. In this case, since all other caches are in
the I state, the number of acknowledgments to wait for is zero. The
cache therefore, immediately updates its local copy of the data with
the new value D and notifies its environment that the request has been
satisfied and transitions to the E state.

On the other hand, Figure~\ref{figure:msi_store_inv} depicts the
scenario when a \emph{write} command is received by a cache and some
other cache is in the S state. In this case, the directory sends
invalidations to all the caches in the S state, and sends a
\emph{DataD2C} message to the requesting cache with the \emph{NumAcks}
field set to the number of sharers, notifying the cache that it needs
to wait for as many invalidate acknowledgments. The other caches
directly communicate with the requesting cache by sending
acknowledgment of the invalidation from the directory. Note that this
is not part of the base German/MSI coherence protocol, where the directory
collects acknowledgments instead. With the extension, the
cache-to-cache communication reduces the amount of processing that
needs to be done in the centralized directory.

\begin{figure}[!t]
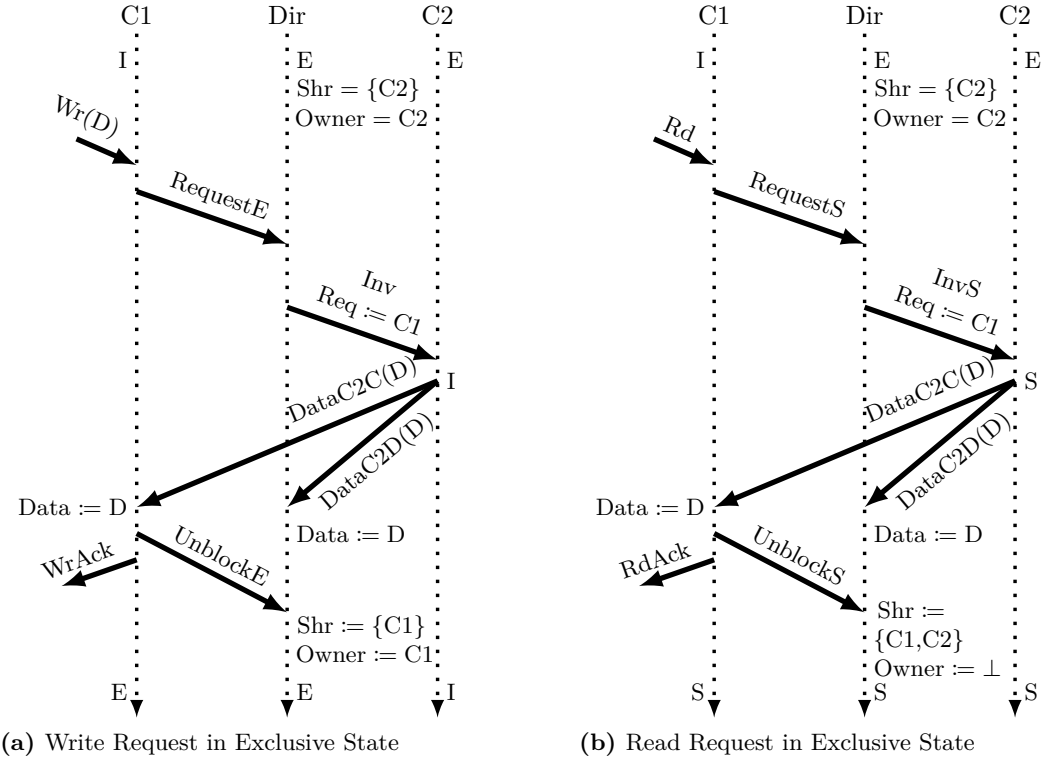

\centering
\subfloat[Write Request in Exclusive State]{
  \input{figures/msi/msi_scenario_header.tex}
\begin{tikzpicture}
  \begin{scope}[x=2cm, y=-.7cm]
    \path
	(0, 0) node[above] {C1}
	(1, 0) node[above] {Dir}
	(2, 0) node[above] {C2}
	(0, 0) edge[lane] (0, 13)
	(1, 0) edge[lane] (1, 13)
	(2, 0) edge[lane] (2, 13)

  	(0, .5) node[left, lane label] {I}
	(1, .5) node[right, lane label] {E}
	(2, .5) node[right, lane label] {E}

    (1, 1.3) node[right, lane label]
    {
      \hspace{-1mm}$\text{Shr} = \{\text{C2}\}$\\
      \hspace{-0.2mm}$\text{Owner} = \text{C2}$
    }
	(-.4, 2)
	edge[message]
	node[message node,pos=0] {Wr(D)}
	(0, 2.5)

	(0, 3)
	edge[message]
	node[message node] {RequestE}
	(1, 4)

	(1, 5.2)
	edge[message]
 	node[message node]
 	{Inv\\$\text{Req}\coloneqq \text{C1}$}
	(2, 6.2)

    (2, 6.6)
    edge[message]
    node[message node, pos=.25] {DataC2C(D)}
    (0, 9)
    (2, 6.6)
    edge[message]
    node[message node, pos=.5, below] {DataC2D(D)}
    (1, 9)

    (2, 6.6) node[right, lane label] {I}

    (1, 9.5) node[right, command node] {$\text{Data} \coloneqq \text{D}$}

    (0, 9) node[left, lane label] {$\text{Data} \coloneqq \text{D}$}

    (0, 9.5)
    edge[message]
    node[message node] {UnblockE}
    (1, 11)

    (0,10)
    edge[message]
    node[message node,pos=0.7] {WrAck}
    (-.5, 10.5)

    (1, 11.5)
    node[right, command node]
    {
      \hspace{-1mm}$\text{Shr} \coloneqq \{\text{C1}\}$\\
      \hspace{-0.1mm}$\text{Owner} \coloneqq \text{C1}$
    }

    (0, 12.5) node[left, lane label] {E}
    (1, 12.5) node[right, lane label] {E}
	(2, 12.5) node[right, lane label] {I}
    ;
  \end{scope}
\end{tikzpicture}
  \label{figure:msi_store_mod}
}
\subfloat[Read Request in Exclusive State]{
  \input{figures/msi/msi_scenario_header.tex}
\begin{tikzpicture}
  \begin{scope}[x=2cm, y=-.7cm]
    \path
	(0, 0) node[above] {C1}
	(1, 0) node[above] {Dir}
	(2, 0) node[above] {C2}
	(0, 0) edge[lane] (0, 13)
	(1, 0) edge[lane] (1, 13)
	(2, 0) edge[lane] (2, 13)

  	(0, .5) node[left, lane label] {I}
	(1, .5) node[right, lane label] {E}
	(2, .5) node[right, lane label] {E}

    (1, 1.3) node[right, lane label]
    {
      \hspace{-1mm}$\text{Shr} = \{\text{C2}\}$\\
      \hspace{-.1mm}$\text{Owner} = \text{C2}$
    }

	(-.4, 2)
	edge[message]
	node[message node,pos=0.3] {Rd}
	(0, 2.5)

	(0, 3)
	edge[message]
	node[message node] {RequestS}
	(1, 4)

	(1, 5.2)
	edge[message]
 	node[message node]
 	{InvS\\$\text{Req}\coloneqq \text{C1}$}
	(2, 6.2)

    (2, 6.6)
    edge[message]
    node[message node, pos=.25] {DataC2C(D)}
    (0, 9)
    (2, 6.6)
    edge[message]
    node[message node, pos=.5, below] {DataC2D(D)}
    (1, 9)

    (2, 6.6) node[right, lane label] {S}

    (1, 9.5) node[right, command node] {$\text{Data} \coloneqq \text{D}$}

    (0, 9) node[left, lane label] {$\text{Data} \coloneqq \text{D}$}

    (0, 9.5)
    edge[message]
    node[message node] {UnblockS}
    (1, 11)

    (0,10)
    edge[message]
    node[message node, pos=0.7] {RdAck}
    (-.5, 10.5)

    (1, 11.5)
    node[right, command node]
    {
      \hspace{-7mm}$\text{Shr} \coloneqq$\\
      \hspace{-5mm}$\{\text{C1,C2}\}$\\
      $\text{Owner} \coloneqq \bot$
    }

    (0, 12.5) node[left, lane label] {S}
    (1, 12.5) node[right, lane label] {S}
	(2, 12.5) node[right, lane label] {S}
    ;
  \end{scope}
\end{tikzpicture}
  \label{figure:msi_load_mod}
}
\caption{Requests in Exclusive State in the German/MSI Protocol}
\label{figure:msi_load_store_mod}
\end{figure}

Figure~\subref*{figure:msi_store_mod} describes the behavior of the
protocol when a cache receives a \emph{write} request and some other
cache in the system is in the E state. The actions are similar to the
case where some other cache is in the S state, except that the cache
already in the E state directly sends its data to the requesting
cache, as well as to the directory. And the requesting cache does not
need to wait for any acknowledgments. Note that this is again an
extension to the base German/MSI protocol, where the data is sent to only
the directory, and the directory forwards the data back to the
requesting cache. Again, this extension reduces the amount of
processing that needs to be handled at the centralized directory.

The scenario when a cache receives a \emph{read} command from the
environment when some other cache in the system is in the E state is
shown in Figure~\subref*{figure:msi_load_mod}. Again, the directory sends
an invalidation to the cache in the E state, which in turn responds by
sending the most up-to-date copy of the data to the directory as well
as the requesting cache. It then downgrades its permissions to the S
state. Both the cache and the directory update their local copies of
the state. The directory in addition adds the requesting cache to set
of sharers.

Figures~\subref*{figure:msi_evict_shared} and~\subref*{figure:msi_evict_mod}
describe the behavior of the protocol in the case where a cache
wishes to relinquish its permissions. This is not a scenario that
occurs in the base German/MSI protocol, but is necessary in a real-world
coherence protocol, where a line of unused data may need to be evicted
to make room for some other data. In the event that the cache in the S
state, it silently evicts the line, without notifying the
directory. This can be done, only because the directory already has
the most up-to-date copy of the data --- recall that the S state only
grants read permissions to the cache, hence it could not have modified
the data. On the other hand if the cache is in the E state, then
it needs to send the most up-to-date copy of the data to the
directory. Therefore it sends a \emph{WriteBack} message to the
directory which contains the most up-to-date copy of the
data. The directory then updates its local copy of the data with this
copy and notes that all caches in the system are in the I state.

\bfpara{Corner-cases in the German/MSI Protocol.}{
We consider a more complex variant of the German cache coherence
protocol to evaluate the techniques we have presented so far, which we
refer to as German/MSI. The main differences from the base German
protocol are: (1) Direct communication between caches is possible in
some cases, (2) A cache in the S state can silently relinquish its
permissions, which can cause the directory to have out-of-date
information about the caches which are in the S state. (3) A cache in
the E state can coordinate with the directory to relinquish their
permissions. We have discussed a list of scenarios typically used when
describing this protocol. These scenarios however, do not describe the
protocol's behavior in several cases induced by concurrency. There are
five such cases that need to be considered in the case of the
German/MSI protocol.

Figure~\ref{figure:msi_race_scenario}, presents the first of these
cases. Initially, cache C1 is in the I state, in contrast, the directory
records that C1 is in state S and is a sharer, due to C1 having
silently relinquished its read permissions at some point in the
past. Now, both caches C1 and C2 receive \emph{write} commands from
their respective environments. Cache C2 sends a \emph{RequestE}
message to the directory, requesting exclusive write permissions. The
directory, under the impression that C1 is in state S, sends an
\emph{Inv} message to it, informing it that C2 has requested exclusive
access and C1 needs to acknowledge that it has relinquished
permissions to C2. Concurrently, cache C1
sends a \emph{RequestE} message to the directory requesting write
permissions as well, which gets delayed. Subsequently, the cache C1
receives an invalidation when it
is in the state IM, which cannot happen in the base German
protocol. The correct behavior for the cache in this situation (shown
by dashed arrows), is to send an \emph{InvAck} message
to the cache C2. The guard, the state variable
updates, as well as the location update is what we have left
unspecified in the case of this particular scenario. As part of the
evaluation, we successfully synthesized the behavior of the German/MSI
protocol in five such corner-case scenarios arising from
concurrency.

\begin{figure}[!t]
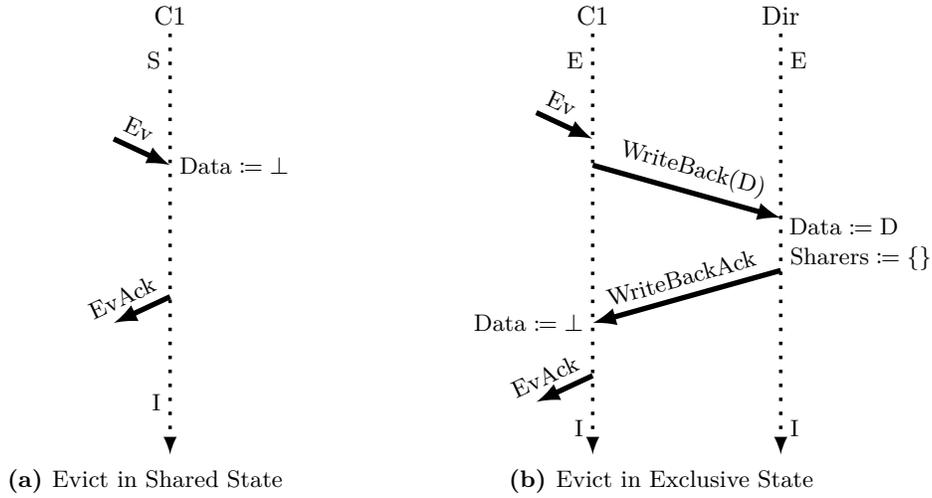

\centering
\subfloat[Evict in Shared State]{
  \input{figures/msi/msi_scenario_header.tex}

\begin{tikzpicture}
  \begin{scope}[x=2.5cm, y=-.7cm]
    \path
	(0, 0) node[above] {C1}
	(0, 0) edge[lane] (0, 8)

  	(0, .5) node[left, lane label] {S}

	(-.3, 2)
	edge[message]
	node[message node, pos=0.3] {Ev}
	(0, 2.5)

    (0, 2.5) node[command node, right] {$\text{Data} \coloneqq \bot$}

    (0, 5)
    edge[message]
    node[message node, pos=0.7] {EvAck}
    (-.3, 5.5)

    (0, 7) node[left, lane label] {I}
    ;
  \end{scope}
\end{tikzpicture}
  \label{figure:msi_evict_shared}
}\qquad
\subfloat[Evict in Exclusive State]{
  \input{figures/msi/msi_scenario_header.tex}
\begin{tikzpicture}
  \begin{scope}[x=2.5cm, y=-.7cm]
    \path
	(0, 0) node[above] {C1}
	(1, 0) node[above] {Dir}
  	(0, .5) node[left, lane label] {E}
	(1, .5) node[right, lane label] {E}
	(0, 0) edge[lane] (0, 8)
	(1, 0) edge[lane] (1, 8)
	(-.3, 1.5)
	edge[message]
	node[message node, pos=0.3] {Ev}
	(0, 2)

	(0, 2.5)
	edge[message]
	node[message node] {WriteBack(D)}
	(1, 3.5)

    (1, 4) node[right, command node]
    {
    \hspace{-4.5mm}$\text{Data} \coloneqq \text{D}$\\
    $\text{Sharers}\coloneqq \{  \}$
    }

	(1, 4.5)
	edge[message]
 	node[message node] {WriteBackAck}
	(0, 5.5)

	(0, 5.5) node[left, command node] {$\text{Data}\coloneqq \bot$}

	(0, 6.5)
    edge[message]
    node[message node, pos=0.7] {EvAck}
    (-.3, 7)

    (1, 7.5) node[right, lane label] {I}
    (0, 7.5) node[left, lane label] {I}
    ;
  \end{scope}
\end{tikzpicture}
  \label{figure:msi_evict_mod}
}
\caption{Evict requests in the German/MSI protocol}
\label{figure:msi_evict}
\end{figure}

\begin{figure}[!t]
  \centering
  \input{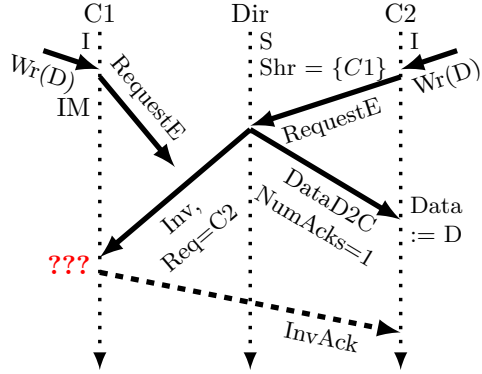}
\begin{tikzpicture}[x=2cm, y=-1cm]
    \path
    (0,0) node [above] {C1}
    (1 ,0) node [above] {Dir}
    (2 ,0) node [above] {C2}

    (0, 0) edge[lane] (0, 4.5)
    (1, 0) edge[lane] (1, 4.5)
    (2, 0) edge[lane] (2, 4.5)

    (0, 0.125) node[left, lane label] {I}
    (1, 0.125) node[right, lane label] {S}
    (1, 0.5) node[right, lane label] {Shr = $\{C1\}$}
    (2, 0.125) node[right, lane label] {I}

    (-.375, 0.25)
    edge[message]
    node[message node,below,pos=0.1] {Wr(D)}
    (0, 0.5)

    (2.375, 0.25)
    edge[message]
    node[message node,below,pos=0.3] {Wr(D)}
    (2, 0.5)

    (0, 0.6)
    edge[intmsg]
    node[message node,above,pos=0.4] {RequestE}
    (0.5, 1.8)

    (0,1) node [left] {IM}

    (2, 0.6)
    edge[message]
    node[message node,below] {RequestE}
    (1, 1.25)

    (1, 1.3)
    edge[message]
    node[message node,below,pos=0.55] {Inv,\\Req=C2}
    (0, 3)

    (1, 1.3)
    edge[message]
    node[message node, below,pos=0.6] {DataD2C\\NumAcks=1}
    (2, 2.5)

    (2, 2.5) node[right, lane label] {Data\\$:=$ D}

    (0, 3.1) node[left] {\red{\textbf{???}}}

    (0, 3.2) edge[tentmessage]
    node[message node, below, pos=0.75] {InvAck}
    (2, 4)
    ;
\end{tikzpicture}
  \caption{Racy Scenario}
  \label{figure:msi_race_scenario}
\end{figure}

\begin{figure}[!t]
\centering
\input{figures/msi/msi_scenario_header.tex}
\begin{tikzpicture}[x=2.5cm, y=-1cm]
    \path
    (0,0) node [above] {C1}
    (1 ,0) node [above] {Dir}
    (2 ,0) node [above] {C2}

    (0, 0) edge[lane] (0, 7)
    (1, 0) edge[lane] (1, 7)
    (2, 0) edge[lane] (2, 7)

    (0, 0.125) node[left, lane label] {E}
    (1, 0.125) node[right, lane label] {E}
    (1, 0.7) node[right, lane label]
    {
      \hspace{-1mm}Shr = $\{\text{C1}\}$\\
      \hspace{-0.1mm}Owner = C1
    }
    (2, 0.125) node[right, lane label] {I}

    (-.3, 0.6)
    edge[message]
    node[message node, pos=0.1] {Ev}
    (0, 1)

    (2.3, 0.6)
    edge[message]
    node[message node, pos=0.1] {Rd}
    (2, 1)

    (2, 1.1)
    edge[message]
    node[message node,below] {RequestS}
    (1, 2)

    (1, 2.1)
    edge[message]
    node[message node, pos=0.7] {Inv}
    (0, 3.1)

    (0,1.1)
    edge[message]
    node[message node, pos=0.35] {WriteBack(D)}
    (1, 3.1)

    (0, 3.1) node[left, lane label] {\red{\textbf{???}}}
    (1, 3.1) node[right, lane label] {\red{\textbf{???}}}

    (1, 3.6) node[right, lane label]
    {
      \hspace{-2.5mm}$\text{Data} \coloneqq \text{D}$\\
      $\text{Owner} \coloneqq \bot$
    }

    (1, 4)
    edge[tentmessage]
    node[message node,below]
    {
      DataD2C(D)\\
      $\text{NumAcks} \coloneqq 0$
    }
    (2, 4.5)

    (0, 4)
    edge[tentmessage]
    node[message node, pos=0.7] {EvAck}
    (-0.3, 4.5)

    (0, 3.5) node[left, lane label] {$\text{Data} \coloneqq \bot$}
    (0, 4.7) node[left, lane label] {I}

    (2, 5) node[right, lane label] {$\text{Data} \coloneqq \text{D}$}

    (2, 5.5)
    edge[message]
    node[message node] {UnblockS}
    (1, 6)

    (2, 5.6)
    edge[message]
    node[message node, pos=0.7] {RdAck}
    (2.3, 6.1)

    (1, 6.3)
    node[right, lane label] {$\text{Shr} \coloneqq \{\text{C2}\}$}

    (0, 6.7) node[left, lane label] {I}
    (1, 6.7) node[right, lane label] {S}
    (2, 6.7) node[right, lane label] {S}
    ;
\end{tikzpicture}
\caption{An interleaving of scenarios which leads to unspecified
  behavior in the MSI/German protocol}
\label{figure:msi_race_scenario_appendix}
\end{figure}

Two of the remaining four scenarios are similar to the one shown in
Figure~\ref{figure:msi_race_scenario}, with the only difference being
that either the \emph{RequestS} message is sent by C1 in response to a
\emph{Read} command from the environment, or that C1 begins in the S
state, and sends a \emph{RequestE} message in response to a
\emph{Write} command.

We now describe the last two corner-cases which arise from
concurrency, in the German/MSI protocol. These are depicted in
Figure~\ref{figure:msi_race_scenario_appendix}. Essentially, the
scenarios shown in
Figures~\subref*{figure:msi_evict_mod}~and~\subref*{figure:msi_load_mod}
interleave, to obtain the situation shown in
Figure~\ref{figure:msi_race_scenario_appendix}. The cache C1 having
sent a \emph{WriteBack} message to the directory is not expecting an
\emph{Inv} message. Similarly, the directory, having sent an
\emph{Inv} to cache C1 is not expecting a \emph{WriteBack} message
from it. The correct way for the processes to behave in this situation
is show by dashed arrows in
Figure~\ref{figure:msi_race_scenario_appendix}. The cache behaves as
if the \emph{Inv} message was a \emph{WritebackAck} message and
notifies its environment of completion. The directory updates its
local copy of the data with the one from the \emph{WriteBack} message,
and then sends this data over to the cache C2, informing it that it
need not wait for any acknowledgment. After this point, both the
cache and directory behaviors know how to interact with each other
as shown in  Figure~\subref*{figure:msi_load_simple}. For completeness,
the way the scenario plays out is shown in
Figure~\ref{figure:msi_race_scenario_appendix} as well.
}

\subsection{Summary of Experimental Results}
Table~\ref{table:experiments_table} summarizes our experimental
findings. All experiments were performed on a Linux desktop, with an
Intel Core i7 CPU running at 3.4 GHz., with 8 GB of memory. The
columns show the name of the benchmark, the number of unknown
functions that were synthesized (\# UF), the size of the search space
for the unknown functions, the number of states in the complete
protocol (\# States), ``symm. red.'' denotes symmetry reduced state
space.  The ``\# Iters.'' column shows the number of algorithm
iterations, while the last two columns show the total amount of time
spent in SMT solving and the end-to-end synthesis time.

The ``German/MSI-$n$'' rows correspond to the
synthesizing the unknown behavior for the German/MSI protocol, with $n$ out
of the five unknown transitions left unspecified. In each case, we
applied the heuristic to obtain minimally permissive guards and biased
the search towards updates
which leave the values of state variables unchanged as far as possible,
except in the case of the Dijkstra benchmark, as mentioned in
Section~\ref{subsec:self_stabilizing_systems}.
Also, note that we ran each benchmark
multiple times with different random seeds to the SMT solver, and report
the worst of the run times in Table~\ref{table:experiments_table}.

\bfpara{Programmer Assistance.}{
  In all cases, the programmer specified the kinds of messages to
  handle in the states where the behavior was unknown.  For example, in
  the case of the German/MSI protocol, the programmer indicated that in
  the IM state on the cache, it needs to handle an invalidation from the
  directory (see Figure~\ref{figure:msi_race_scenario}). In general, the
  programmer specified \emph{what} needs to be handled, but not the
  \emph{how}. This was crucial to getting our approach to scale.
}

\bfpara{Overhead of Decision Procedures.}{
  We observe from Table~\ref{table:experiments_table} that for the longer
  running benchmarks, the run time is dominated by SMT solving. In all
  of these cases, a very large fraction of the constraints
  asserted into the SMT solver are constraints to implement heuristics
  which are specifically aimed at guiding the SMT solver, and reducing
  the impact of non-deterministic choices made by the solver.
  Specialized decision procedures that
  handle these constraints at an algorithmic level~\cite{bjorner-14} can
  greatly speed up the synthesis procedure.% \footnote{We are awaiting a
    % stable release of $\nu$Z to evaluate our techniques using an optimizing
    % SMT solver. The current version of $\nu$Z, available from the project's
    % source code repositories lacks documentation and sees
    % regular commit activity.}
}

\bfpara{Synthesizing Symbolic Expressions.} {
  The interpretations returned by the SMT solver are in the form of tables,
  which specify the output of the unknown function on specific inputs.
  We mentioned that if a symbolic expression is required we can pass
  this output to a \sygus solver, which will then return a symbolic
  expression. We were able to synthesize compact expressions in all
  cases using the enumerative \sygus
  solver~\cite{udupa-fmcad-sygus}. Further, although the
  interpretations are only guaranteed to be correct for the finite
  instance of the protocol, the symbolic expressions generated by the
  \sygus solver were \emph{parametric}. We found that they were
  general enough to handle larger instances of protocol as well.
}

\begin{table}[!t]
\centering
\begin{tabular*}{\linewidth}{@{\extracolsep{\fill}}lllllll}\hlx{hvhv}
  \multicolumn{1}{c}{\multirow{2}{*}{Benchmark}} &
  \multicolumn{1}{c}{\multirow{2}{*}{\# UF}} &
  \multicolumn{1}{c}{Search} &
  \multicolumn{1}{c}{\multirow{2}{*}{\# States}} &
  \multicolumn{1}{c}{\multirow{2}{*}{\# Iters.}} &
  \multicolumn{1}{c}{SMT} &
  \multicolumn{1}{c}{Total} \\
  & & \multicolumn{1}{c}{Space} & & & \multicolumn{1}{c}{Time} &
  \multicolumn{1}{c}{Time} \\\hlx{vhv}
Peterson & 3 & $2^{36}$ & 60 & 14 & 97ms & 130ms\\
Dijkstra & 6 & $2^{192}$ & \textasciitilde 2000 & 30 & 27s & 64s\\
% Dijkstra-13 & 6 & $2^{192}$ & \textasciitilde 2000 & 184 & 181s & 658s\\
% Dijkstra-15 & 6 & $2^{192}$ & \textasciitilde 2000 & 127 & 34s & 338s\\
German/MSI-2 & 16 & \textasciitilde$2^{4700}$ & \textasciitilde 20000
(symm. red.) & 217 & 31s & 298s\\
German/MSI-4 & 28 & \textasciitilde$2^{7614}$ & \textasciitilde 20000
(symm. red.) & 419 & 898s & 1545s\\
German/MSI-5 & 34 & \textasciitilde$2^{9000}$ & \textasciitilde 20000
(symm. red.) & 525 & 2261s & 3410s\\
\hlx{hvh}
\end{tabular*}
\caption{Experimental Results}
\label{table:experiments_table}
\end{table}

%%% Local Variables:
%%% TeX-master: "cav15"
%%% End:

\section{Conclusions}
\label{sec:conclusions}
We have presented an algorithm to complete symmetric distributed
protocols specified as \textsc{esm} sketches, such that they satisfy
the given safety and liveness properties. A prototype implementation,
which included a custom model checker, successfully
synthesized non-trivial portions of Peterson's mutual exclusion
protocol, Dijkstra's self-stabilizing system, and the German/MSI cache
coherence protocol. We show that programmer assistance in the form of
\emph{what} needs to be handled is crucial to the scalability of the
approach. Scalability is currently limited by the constraints that
require candidate interpretations to be intuitively simple. Note that
these heuristics were necessary to reduce the dependence of the run
time of our algorithm on the non-deterministic choices made by the SMT
solver. As part of future work, we plan to investigate decision
procedures that implement these heuristics at an algorithmic level,
rather than using constraints which are treated the same as the
correctness constraints.

\bibliographystyle{plainnat}
\setlength{\bibsep}{3pt}
\bibliography{main}
\end{spacing}
\end{document}